\documentclass{aa} 
\usepackage{psfig}
\usepackage{natbib}
\bibpunct{(}{)}{,}{a}{}{,} 
\usepackage{graphicx}

\def\la{\;
\raise0.3ex\hbox{$<$\kern-0.75em\raise-1.1ex\hbox{$\sim$}}\; }
\def\ga{\;
\raise0.3ex\hbox{$>$\kern-0.75em\raise-1.1ex\hbox{$\sim$}}\; }

\newcommand{\kms}{km~s$^{-1}\,$}

\newcommand{\cm}{cm$^{-2}\,$}

\newcommand{\etal}{{et al.}}
\newcommand{\zem}{$z_{\rm em}$\,}
\newcommand{\zabs}{$z_{\rm abs}$\,}

\begin{document}

\title{Fluctuations of the intergalactic ionization field at redshift $z \sim 2$
}

\author{
I. I. Agafonova\inst{1,2}
\and
S. A. Levshakov\inst{1,2,3}
\and
D. Reimers\inst{1}
\and
H.-J. Hagen\inst{1}
\and
D. Tytler\inst{4}
}
\institute{
Hamburger Sternwarte, Universit\"at Hamburg,
Gojenbergsweg 112, D-21029 Hamburg, Germany
\and
Ioffe Physical-Technical Institute, 
Polytekhnicheskaya Str. 26, 194021 St.~Petersburg, Russia\\
\email{ira@astro.ioffe.rssi.ru} 
\and
St.~Petersburg Electrotechnical University `LETI', Prof. Popov Str. 5, 
197376 St.~Petersburg, Russia
\and
Center for Astrophysics and Space Science, University of California, San Diego, CA 92093-0424, USA
}
\date{Received 00  ; Accepted 00}
\abstract
{}
{To probe the spectral energy distribution (SED) of the ionizing background radiation at $z \la 2$
and to specify the sources contributing to the intergalactic radiation field.
} 
{The spectrum of a bright quasar \object{HS~1103+6416} (\zem\ = 2.19) contains five successive metal-line
absorption systems at \zabs\ = 1.1923, 1.7193, 1.8873, 1.8916, and 1.9410.
The systems are optically thin and reveal multiple lines of different metal ions with
the ionization potentials lying in the extreme ultraviolet (EUV) range ($\sim$1 Ryd to $\sim$0.2 keV).
For each system, the EUV SED  of the underlying ionization field 
is reconstructed by means of a special technique developed for solving the inverse problem
in spectroscopy. For the \zabs\ = 1.8916 system, the analysis also involves the \ion{He}{i} resonance lines
of the Lyman series and the \ion{He}{i}~$\lambda504$ \AA\
continuum, which are seen for the first time in any cosmic object except the Sun.  
}
{From one system to another, the SED of the ionizing continuum changes significantly, indicating that 
the intergalactic ionization field at $z \la 2$
fluctuates at the scale of at least $\Delta_z \sim 0.004$. This is consistent with $\Delta_z \la 0.01$ estimated
from \ion{He}{ii} and \ion{H}{i} Lyman-$\alpha$ forest measurements between the redshifts 2 and 3.
A radiation intensity break by approximately an order of magnitude at $E = 4$ Ryd in SEDs 
restored for the \zabs\ = 1.1923, 1.8873, 1.8916, and 1.9410 systems
points to quasars as the main sources of the ionizing radiation. 
The SED variability is mostly caused by a small number of objects contributing
at any given redshift to the ionizing background;
at scales $\Delta_z \ga 0.05$, the influence of local radiation sources becomes significant.
A remarkable SED restored for the \zabs\ = 1.7193 system, with a sharp break shifted to $E \sim 3.5$ Ryd and
a subsequent intensity decrease by $\sim$1.5 dex, indicates a source with comparable inputs of both 
hard (active galactic nuclei, AGN) and soft (stellar) radiation components.
Such a continuum can be emitted by (ultra) luminous infrared galaxies, many of which reveal both a strong AGN
activity and intense star formation in the circumnuclear regions.}
{}
\authorrunning{I. I. Agafonova \etal\ }
\titlerunning{Fluctuating intergalactic ionization field at  $z \sim 2$}

\maketitle

\section{Introduction}
\label{sect-1}

Lines of different ions observed in spectra of
extragalactic sources indicate that the intergalactic medium (IGM) is kept ionized.
Fractions of individual ions are determined by the spectral energy distribution (SED)
of the ionizing continuum; hence, 
a particular shape of the SED must be known
to transform the measured line intensities into absolute element abundances.
Another reason to study the SED is that it provides information about the inputs
of hard (active galactic nuclei/quasars, AGNs/QSOs) and soft (star-forming galaxies) 
radiation emitters. Thus it can help to specify the relative contributions of different 
sources to the integrated ionizing continuum.

The spectral range considered in the present paper is the extreme ultraviolet (EUV), defined 
as energies between the hydrogen 
ionization potential at 1 Ryd (13.6 eV) and  $\sim$0.2 keV (15 Ryd), where the soft X-ray band begins.
The SED in the EUV range is still poorly known
since direct observations are difficult due to several technical and physical limitations. 
However, the EUV SED can be probed indirectly by using spectral lines of different ions observed in the
absorption spectra of distant quasars, i.e., by solving the inverse problem when the shape of the underlying
ionizing continuum is restored from the measured column densities (which are proportional to the ion fractions). 
The simplest (coarse) SED reconstruction involving only the ratio of the radiation intensities
at 4 and 1 Ryd (so-called hardness of the spectrum) can be performed on the basis of  
the column densities of \ion{He}{ii} and \ion{H}{i} lines detected in the Ly-$\alpha$ forest.
Up to now, such measurements have been carried out for the \ion{H}{i} and \ion{He}{ii} lines 
between redshifts $z \sim 2$ and $\sim 3$ along 
two different lines of sight (toward \object{HE~2347--4342} and \object{HS~1700+6416}),
and it was found that the parameter $\eta = $\ion{He}{ii}/\ion{H}{i} fluctuates
from one spatial coordinate to another
with a characteristic small scale of $\Delta_z \la 0.01$ 
(Reimers \etal\ 1997; 
Zheng \etal\ 2004; Shull \etal\ 2004, 2010;
Fechner \etal\ 2006, 2007).
This was interpreted as a result of a variable hardness of the ionizing radiation 
caused by diversity
of the radiation sources coupled with radiation transfer effects in
the non-uniform IGM.
The discovery of foreground quasars located
transverse to the line of sight just at the positions where the \ion{He}{ii}/\ion{H}{i} ratios
show their lowest values
supports the assumption that the metagalactic ionizing radiation is strongly affected by local sources
(Jakobsen \etal\ 2003; Worseck \etal\ 2007).

A finer reconstruction of the SED 
requires the use of metal absorption lines 
(\ion{C}{ii}--\ion{C}{iv}, \ion{Si}{ii}--\ion{Si}{iv},
\ion{N}{ii}--\ion{N}{v}, \ion{O}{i}--\ion{O}{vi}, etc.),
ionization potentials of which sample the EUV range of the ionizing continuum with a quite small step. 
An important fact is that at $z \la 2$ the metal lines
are the sole means to probe the continuum shape  since with modern facilities  
the \ion{He}{ii} Ly-$\alpha$ forest can be studied only at $z > 2$.
Different variants of the SED reconstruction procedure and its application to 
the analysis of specific quasar absorption-line systems are described in
Agafonova \etal\ (2005), Levshakov \etal\ (2008, 2009), and Fechner (2011).

In the present work we restore the SED of the ionizing background
for absorption-line systems detected toward a bright quasar \object{HS~1103+6416} 
with \zem\ = 2.19 (Reimers \etal\ 1995).
To reconstruct the SED from metal lines, one generally needs special
absorbers: optically thin systems with lines of successive ions of the same element
(e.g., \ion{C}{ii}--\ion{C}{iii}--\ion{C}{iv}, \ion{Si}{ii}--\ion{Si}{iii}--\ion{Si}{iv}).
However, they are not often found in quasar spectra.
Thus the quasar \object{HS~1103+6416} is one of the most favorable objects 
since its spectrum contains just several  systems suitable for the SED reconstruction 
and detached from each other by
$\Delta_z \sim 0.004 - 0.17$, i.e., by the redshift interval comparable with that 
in the \ion{He}{ii} Ly-$\alpha$ forest studies.
An additional advantage is that the observational data are available in a broad
wavelength range from 1400 \AA\ to 5700 \AA, which provides a large variety
of ionic transitions and, hence, improves the accuracy of the results.

The paper is structured as follows: the computational method used to reconstruct 
the continuum shape
of the ionization field is described in Sect.~\ref{sect-2}.
Section~\ref{sect-3} deals with the observational data with attention
to the consistency of the wavelength calibration
in different parts of the spectrum. In Sect.~\ref{sect-4},
individual absorption systems are analyzed and the results
obtained are discussed in Sect.~\ref{sect-5}.
Our conclusions are given in Sect.~\ref{sect-6}.

\section{Computational method}
\label{sect-2}

The computational procedure used to reconstruct the SED of the background ionizing radiation
from metal lines was considered in detail in Agafonova \etal\ (2005). Here we describe it
briefly to explain the basic principles of the analysis.

\begin{figure}
\vspace{0.0cm}
\hspace{-2.2cm}\psfig{figure=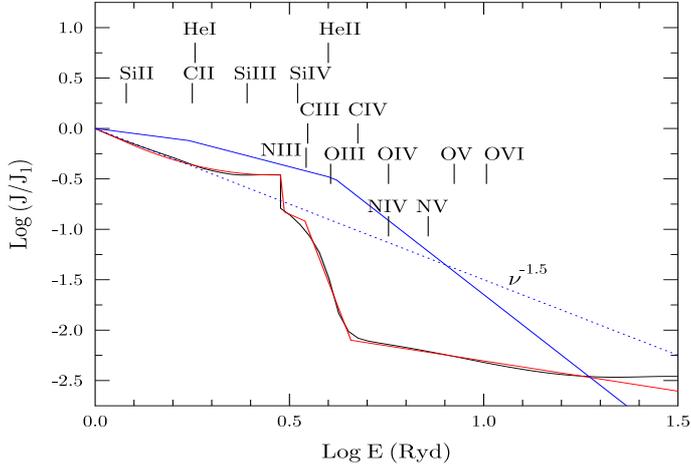,height=15cm,width=12.8cm}
\vspace{-4.0cm}
\caption[]{
Continuum shapes used to approximate the SED in the EUV (1~Ryd -- 0.2~keV) range:
a simple power law (blue dotted line), an AGN continuum by Mathews \& Ferland (1987) (blue line),
an ionizing spectrum of the metagalactic radiation field at redshift $z \sim 2$
by Haardt \& Madau (1996) (black line), and its approximation by a broken power law (red line). 
All the SEDs are normalized so that $J_\nu(h\nu =$ 1 Ryd) = 1.
The positions of ionization thresholds of different
ions are indicated by tick marks.
}
\label{fg1}
\end{figure}

The true shape of the incident radiation can be restored from absorption systems
that are optically thin in the Lyman continuum both in \ion{H}{i} and \ion{He}{ii}, i.e.,
from those with column densities $N$(\ion{H}{i}) $< 2\times10^{17}$ \cm\ and
$N$(\ion{He}{ii}) $< 10^{18}$ \cm.
Another requirement is that the absorbing gas should be in the thermal and ionization equilibrium,
i.e., its ionization structure is
determined entirely by the SED of the ionizing radiation and the ionization parameter 
$U \equiv n_{\rm ph}/n_{\rm gas}$, where $n_{\rm ph}$ and $n_{\rm gas}$ are, respectively, the
local densities of the ionizing photons and of the gas. 
Figure~\ref{fg1} shows continuum shapes commonly used to approximate the EUV range: a simple power law,
an AGN spectrum of Mathews \& Ferland (1987), and an SED of the intergalactic ionization field at $z < 2$
calculated by, e.g., Haardt \& Madau (1996) or Fardal \etal\ (1998).
In this figure, tick marks indicate the ionization
thresholds of ions usually observed in quasar absorption spectra. The column densities of these ions
are proportional to their fractions $\Upsilon_i$, 
which in turn are determined by the SED above the corresponding ionization thresholds.
Thus, having the column densities of ions with ionization potentials sampling some energy interval 
of the ionizing continuum, we can try to solve the inverse
problem: namely, to reconstruct the SED in this interval.
The procedure is iterative and consists of the following steps.

Firstly, the shape of an SED should be parameterized,
i.e., described by a set of parameters (also called factors).
As seen from Fig.~\ref{fg1}, in the EUV range the intensity $J_\nu$ can be well fitted
by a simple/broken power law, i.e., by
a set of $n$ power-law indices $\alpha_i$ ($J_\nu \propto \nu^{-\alpha_i}$)
and breaking points $\nu_i$, $i = 1, \ldots, n$.
The number and definition of factors depend on how many lines of different ions
are available in the absorption system: it is clear that with only a couple of lines
the continuum shape cannot be reconstructed in great detail, i.e., the fewer the lines the fewer
the factors.
With some initial guesses for the adopted factors, a trial SED is computed 
and inserted into the
photoionization code CLOUDY (ver.07.02.01, last described by Ferland \etal 1998).
The ionization parameter $U$ at this stage
of the analysis is assumed constant within the absorber and is commonly
estimated by ratio(s) of column densities of two or more subsequent ions of
the same element
(\ion{Si}{ii}--\ion{Si}{iii}--\ion{Si}{iv}, or \ion{C}{ii}--\ion{C}{iii}--\ion{C}{iv}).
If for this value of $U$  the trial SED cannot reproduce the column densities of all ions 
observed in the system
or some other peculiarities arise (like abundance pattern 
well beyond observational and theoretical constraints), then it is 
considered as inadequate and has to be adjusted.
To evaluate 
the goodness of a trial continuum shape, we need to choose some quantitative measure (`response'). 
The choice is performed rather heuristically and accounts for both the information 
obtained with the
tried SED and for any other a priori available information.
If, for example, the trial SED gives a lower than the observed ratio of
\ion{Si}{ii}/\ion{Si}{iv} at $U$ corresponding to the
observed \ion{C}{ii}/\ion{C}{iv} ratio, then the factor values should be varied in a way
to produce SEDs that will give a higher \ion{Si}{ii}/\ion{Si}{iv}.
In addition, a trial SED can possibly reproduce well the observed
column densities of silicon and carbon ions, 
but only at the expense of an extremely high relative overabundance of silicon to carbon,
[Si/C] $> 0.3$, which is
neither predicted for nor measured in intergalactic absorbers.
In this case the adjustment should go toward the continuum
shapes reducing the [Si/C] value to the acceptable level of [Si/C] $\la 0.3$.
When the SED is parameterized by a single factor (simple power law $J_\nu \propto \nu^{-\alpha}$), then the
adjustment can easily be performed by a trial-and-error procedure.
In case of multiple factors, the direction (in the factor space) toward the desired response
is found by means of a special randomization technique for the factor values based on the experimental
design methodology (details are given in Agafonova \etal\ 2005). 
The adjustment of an initial SED occurs by
stepwise movement along this direction until the SED
that produces the required response is found.
 
Now with the found SED as an input, 
CLOUDY is used to calculate the ionization curves for all ions
detected in the absorption system.
The ionization curves are then inserted in the  Monte Carlo Inversion (MCI) code, which is
a solver for the inverse problem in spectroscopy based on a model of line formation
in clumpy stochastic media, i.e., in the gas with fluctuating density and velocity
fields (full description is given in Levshakov \etal\ 2000).
The MCI reconstructs the distributions of the gas
density (and hence of the ionization parameter)
and the velocity along the line of sight and estimates
the element abundances (assumed to be constant within the absorber) by fitting the synthetic
line profiles to the observed ones, i.e., it works 
directly with the intensities at every point within the line profiles.
Compared with the estimates based solely on column densities, the abundances obtained by the MCI are more accurate 
and robust, especially in case of
complex line profiles, i.e., when significant density and velocity gradients
are expected (see Eq.(15) in Levshakov \etal\ 2000).
Another advantage of the MCI is the treatment of blends: supposing that all lines arise in the same gas,
we restore the underlying gas density and velocity distributions using firstly the lines with clear
profiles and then reconstructing the profiles of blended lines on the basis of these distributions and ionization
curves for the corresponding ions. 
The important point here is that this approach also allows us to check
(and to correct) the eventual shifts of absorption lines caused by errors in wavelength calibration
(Levshakov \etal\ 2009). 
If with a newly found SED we can self-consistently describe the profiles of all observed lines,
i.e., obtain $\chi^2 \la 1$ (per degree of freedom) for individual lines along with
acceptable element abundances, then calculations are
finished. Otherwise, the next iteration of the SED adjustment is performed, 
eventually with a different response function.

\section{Observational data}
\label{sect-3}

Details of the observations and data reduction are given in K\"ohler \etal\ (1999).
Here we repeat only what is relevant for the present study.
A high resolution (FWHM $\sim 8$ \kms) optical spectrum of \object{HS~1103+6416}
was obtained in the wavelength coverage of 3180--5800 \AA\ with the HIRES spectrograph (Vogt \etal\ 1994) 
on the Keck 10 m telescope.
Low-resolution spectra in the UV range were taken with the FOS and the GHRS onboard the
Hubble Space Telescope (HST).
Depending on the wavelength range, the spectral
resolution was 0.77 \AA\ (grating G140L, $\Delta \lambda = 1415-1700$ \AA),
1.44 \AA\ (G190H, $\Delta \lambda = 1572-2311$ \AA),
and 2 \AA\ (G270H, $\Delta \lambda = 2223-3277$ \AA), which is
equivalent to velocity resolutions (FWHM) of
136--163 \kms, 186--275 \kms, and 183--270 \kms, respectively.

Since the data were obtained with different instruments, offsets in the wavelength
calibration between the corresponding parts of the spectrum are possible.
In our analysis of an absorption-line system,
all transitions are fitted with the same model of the underlying density-velocity field and
artificial (calibration-induced) velocity shifts between the absorption lines can negatively affect the 
final result.
To reveal the eventual offsets, we used several absorption systems with strong
hydrogen and metal lines throughout the whole available spectral range. 
The velocity structure was
estimated by fitting the line profiles from the high-resolution Keck spectra to a
number of Gaussian component. The synthetic profiles of low-resolution lines
were then calculated by convolving the obtained velocity structure with a FWHM at
the position of a given line and adjusting the column density to comply with the observed
line intensities. The calibration shifts were evaluated by
cross-correlation of synthetic and observed profiles.

\begin{figure*}
\vspace{0.0cm}
\hspace{-0.5cm}\psfig{figure=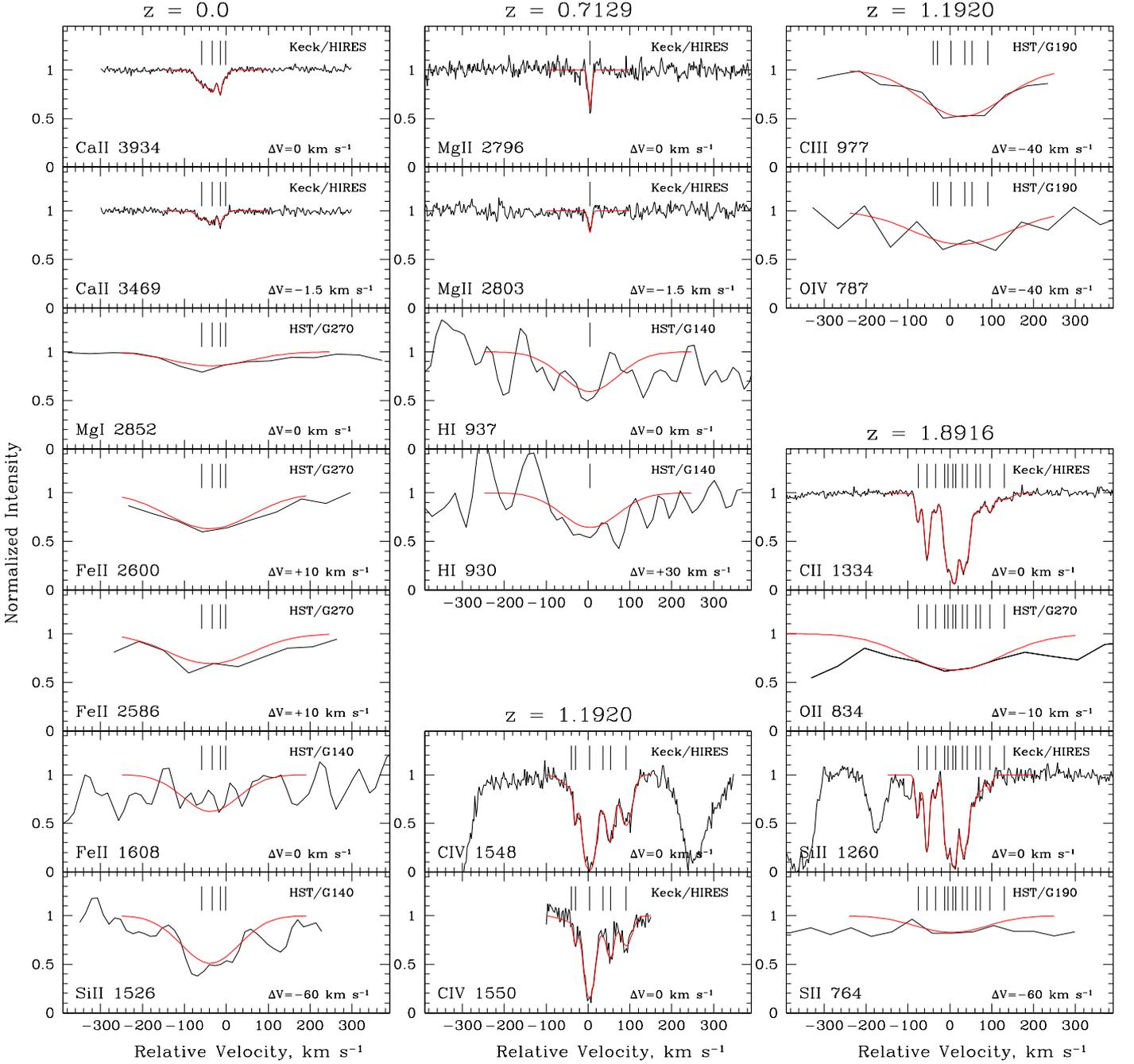,height=18cm,width=19.0cm}
\vspace{0.0cm}
\caption[]{
Calibration-induced shifts of the observed profiles illustrated by spectral lines
from different absorption systems. The velocity structure
is estimated on the basis of lines from the high-resolution (Keck/HIRES) part of the
\object{HS~1103+6416}
spectrum and then applied to calculate the profiles from low-resolution (HST) parts. 
The synthetic profiles are shown by red,
the vertical ticks mark positions of sub-components.
$\Delta V$ indicates the velocity shift applied to the observed profile to align it 
with a reference line from the Keck spectrum.
}
\label{fg2}
\end{figure*}

Some examples are shown in Fig.~\ref{fg2}. For the system at $z = 0$ (probably a DLA),
the velocity structure was estimated on the basis of the
doublet \ion{Ca}{ii} $\lambda\lambda 3439, 3969$ \AA\
in the Keck spectrum and then transferred to other ions assuming purely turbulent broadening
(i.e., the Doppler $b$-parameters for all ions were identical). 
For the systems at $z = 1.1923$ and 1.8916,
the lines of ions with close ionization potentials and close atomic weights 
(e.g., \ion{C}{iv} and \ion{O}{iv}, \ion{Si}{ii} and \ion{S}{ii}) were selected. 
For the system at $z = 0.7129$, a weak
\ion{H}{i} Lyman limit is present in the G140 part of the spectrum, which allows us to estimate the \ion{H}{i}
column density as $N$(\ion{H}{i}) $= (1.0-1.5)\times10^{16}$ \cm, i.e., the  system is optically thin. 
The $b$-parameter for \ion{H}{i}
lines in this system was calculated from the $b$-parameter of \ion{Mg}{ii} doublet in the Keck spectrum assuming
a purely thermal broadening.

It is seen from Fig.~\ref{fg2} that the calibration in the G270 part either complies 
with that in the Keck spectrum (e.g., \ion{Mg}{i} $\lambda 2852$ \AA\ at $z = 0$) or
demonstrates a negative offset of $\sim -10$ \kms (\ion{Fe}{ii} $\lambda\lambda 2600, 2586$ \AA\ at $z = 0$,
\ion{O}{ii} $\lambda 834$ \AA\ at $z = 1.8916$). 
In G190, the offsets relative to the Keck spectrum are much larger ($\sim 40$ \kms) 
and change only slowly with wavelength (\ion{S}{ii} $\lambda 764$ \AA\ at
$z = 1.8916$, \ion{O}{iv} $\lambda 787$ \AA\ at $z = 1.1923$). 
Contrary to this, the calibration in G140 is quite irregular 
and lines detached in the spectrum only by a few angstroms can be either coherent with the 
Keck spectrum (\ion{Fe}{ii} $\lambda 1608$ \AA\
at $z = 0$, \ion{H}{i} $\lambda 937$ \AA\ at $z = 0.7129$)
or significantly shifted (\ion{Si}{ii} $\lambda 1526$ \AA\ at $z = 0$, \ion{H}{i} $\lambda 930$ \AA\ at $z = 0.7129$).

The absorption lines in the Keck spectrum are also not free from calibration errors.
For example, in the system at $z = 0$, the
\ion{Ca}{ii} $\lambda 3969$ \AA\ line is offset by 1.5 \kms\ relative to
\ion{Ca}{ii} $\lambda 3934$ \AA, while
in the system at $z = 0.7129$, the \ion{Mg}{ii} $\lambda 2803$ \AA\ line 
is shifted by $-1.3$ \kms\ relative
to \ion{Mg}{ii} $\lambda 2796$ \AA. 
The calibration of lines of the same ion can be easily checked by a simple cross-correlation
of the line profiles. To reveal possible calibration shifts between lines of different ions, 
several trial MCI runs are required. 
We found that the \ion{Si}{iv} $\lambda 1393$ \AA\ line in the $z = 1.8916$ system 
was offset by 2.0 \kms\ relative
to both low (\ion{Si}{ii}, \ion{C}{ii}) and high ionization lines (\ion{C}{iv}), 
i.e., its shift was caused not by a
kinematical segregation but by miscalibration of the wavelength scale.
In general, we confirm the results of Griest \etal\ (2010) 
that calibration errors in Keck spectra
are quite common and can reach a value of $\sim 0.5$ pixel size ($\sim 2$ \kms).

For calculations described below, all revealed calibration shifts were corrected, 
i.e., positions of lines in each absorption system  
were set according to a coherent wavelength calibration.

\section{Analysis of individual systems}
\label{sect-4}

All calculations throughout the paper are performed with laboratory wavelengths 
and oscillator strengths taken from Morton (2003)
for $\lambda > 912$ \AA, except for the ion \ion{Si}{iii}, and from 
Verner \etal\ (1994) for $\lambda < 912$ \AA. 
For \ion{Si}{iii}, the adopted laboratory wavelength is 
1206.51 \AA\ and not 1206.50 \AA\ 
as Morton's (for explanation, see Sect.~2 in Levshakov \etal\ 2009). 
The Lyman series limit for \ion{He}{i},
$\lambda_c = 504.2593$ \AA, is taken from the database of NIST. 
We use
element abundances of the solar photosphere as given 
in Asplund \etal\ (2009) and Grevesse \etal\ (2010).
The abundance 
ratio [X/Y] means $\log(N_{\rm X}/N_{\rm Y}) - \log(N_{\rm X}/N_{\rm Y})_\odot$,
while [X] = $\log(N_{\rm X}) - \log(N_{\rm X})_\odot$.

\subsection{Absorption system at $z = 1.9410$}
\label{sub-sect-4-1}

\begin{figure*}
\vspace{0.0cm}
\hspace{-0.5cm}\psfig{figure=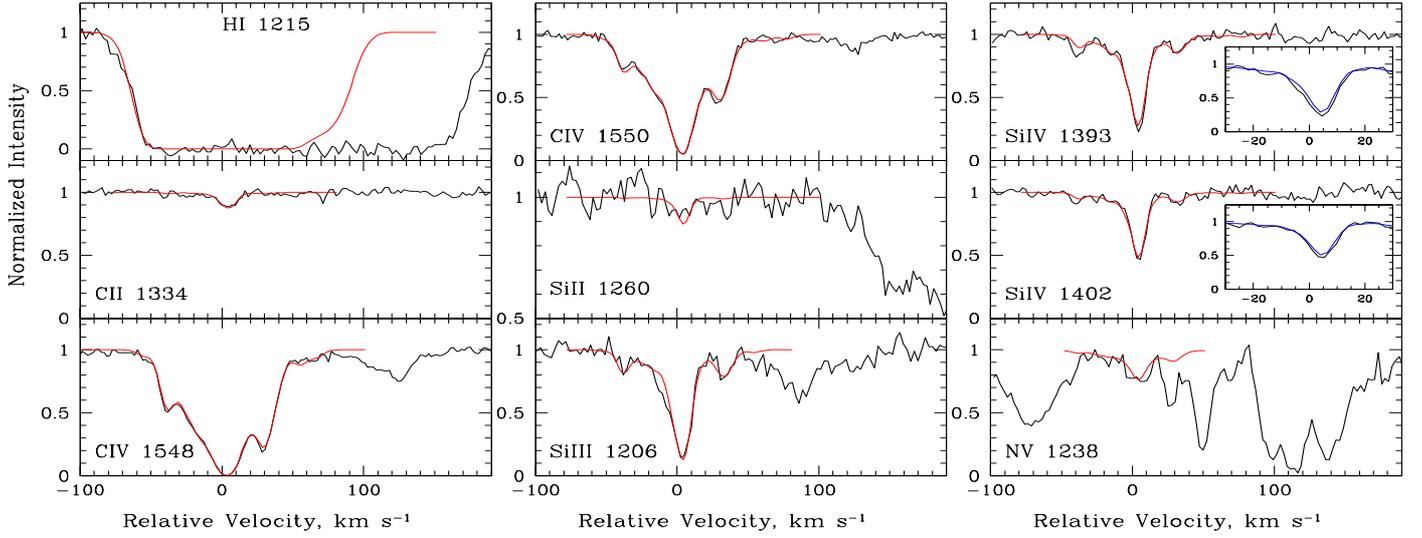,height=16cm,width=19.0cm}
\vspace{-4.5cm}
\caption[]{
High-resolution hydrogen and metal absorption lines associated with the \zabs\ = 1.9410 
system toward
\object{HS~1103+6416} (black curves). 
The synthetic profiles calculated with the UV SEDs shown by the shadowed area in Fig.~\ref{fg5}
are plotted by red curves.
The zero radial velocity is fixed at $z = 1.9410$. 
The inserted figures in \ion{Si}{iv} panels display the
synthetic spectra (blue curves) calculated with the
first-guess SED (red dotted line in Fig.~\ref{fg5}).  
}
\label{fg3}
\end{figure*}

\begin{figure*}
\vspace{0.0cm}
\hspace{-0.5cm}\psfig{figure=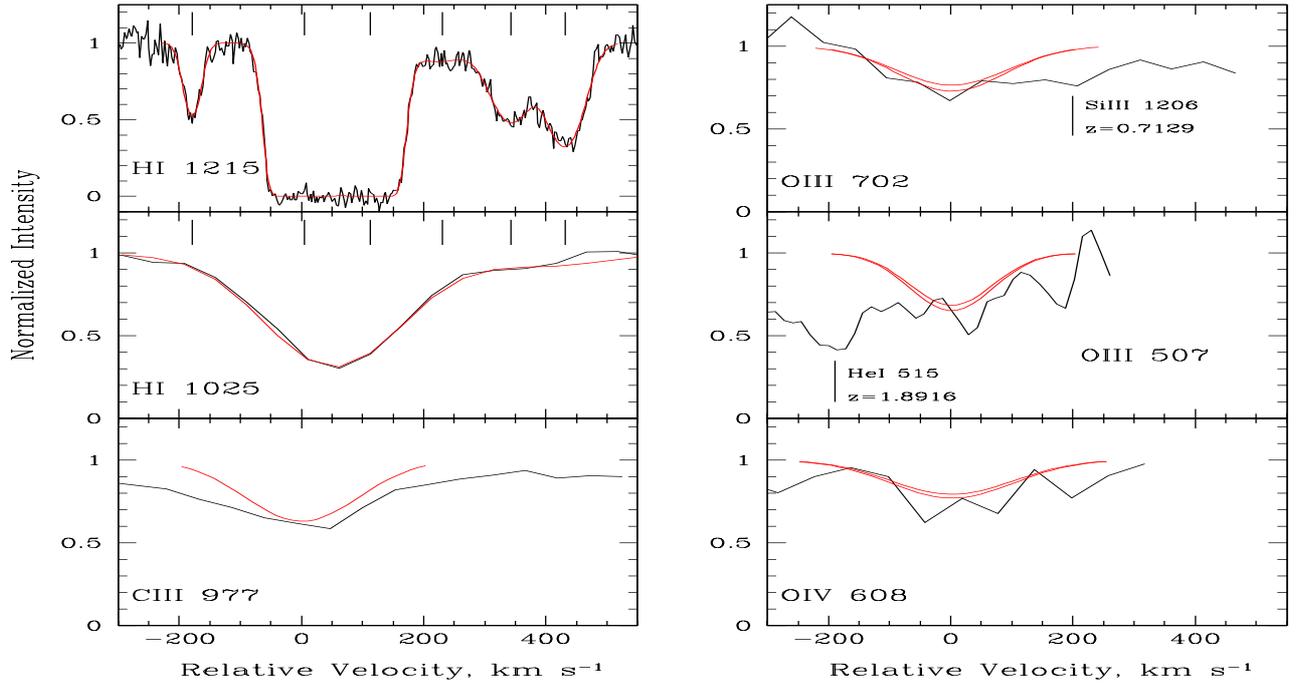,height=15cm,width=28.0cm}
\vspace{-2.5cm}
\caption[]{
Low-resolution lines (HST) from the \zabs\ = 1.9410 system. 
The vertical ticks in the \ion{H}{i} panels indicate the positions of individual Voigt components. 
The profiles
of metal ions are calculated using the velocity-density distributions restored on the basis of high-resolution
lines shown in Fig.~\ref{fg3}. 
The red synthetic profiles in the \ion{O}{iii} and \ion{O}{iv} panels are
calculated for two oxygen abundances [O/C] = 0.2 and 0.3 (see text for details).
}
\label{fg4}
\end{figure*}

The high-resolution (Keck) spectrum of \object{HS~1103+6416} contains the \ion{H}{i} Ly-$\alpha$ line and
lines of subsequent ions of both carbon, 
\ion{C}{ii} $\lambda 1334$ \AA,  \ion{C}{iv} $\lambda\lambda 1548, 1550$ \AA,
and silicon, \ion{Si}{ii} $\lambda 1260$ \AA, \ion{Si}{iii} $\lambda1206$ \AA,
\ion{Si}{iv} $\lambda\lambda1393, 1402$ \AA\ (Fig.~\ref{fg3}, observed profiles are shown by black).
There are absorption features at the expected positions of the 
\ion{N}{v} $\lambda\lambda1238, 1242$ \AA\ lines,
but their profiles are inconsistent with each other, probably because of 
blending with some forest lines.
In any case, they can be used to set an upper limit on the nitrogen abundance.
Clearly present
in the UV part of the spectrum are strong lines \ion{O}{iii} $\lambda\lambda 832, 702, 507$ \AA,
\ion{O}{iv} $\lambda\lambda 608, 553, 554$ \AA, and \ion{C}{iii} $\lambda 977$ \AA\ (Fig.~\ref{fg4}).

The column densities for metal ions from the optical (Keck) spectrum that can be calculated
by a simple Voigt profile fitting  are given in Table~\ref{tbl-1}. 
The column densities for lines from the UV part cannot be 
estimated a priori
due to low resolutions (FWHM $> 200$ \kms) and, as a consequence, possible blending of many 
lines within one profile,
but they can be calculated at the next stages of the SED reconstruction employing the gas density 
and velocity distributions derived by the MCI procedure.

The ion fractions depend strongly on the input metal abundances, so that for accurate analysis we need to know
the \ion{H}{i} content as well.
In Fig.~\ref{fg3}, a neighboring system, which is
seen mainly in the \ion{C}{iv} absorption, is clearly visible at radial velocity $v \simeq 115$ \kms,
i.e., the observed profile of \ion{H}{i} Ly-$\alpha$ is formed by a superposition
of two saturated lines.
Using  \ion{H}{i} lines from the low-resolution (HST) spectrum of \object{HS~1103+6416},  we  
deconvolve this \ion{H}{i} profile into two sub-components, with
$N$(\ion{H}{i}) = $(2.5 - 3.0)\times10^{15}$ \cm\ ($b = 33.1$ \kms) for the $z = 1.9410$ system
and $N$(\ion{H}{i}) = $(0.8 - 1.1)\times10^{15}$ \cm\ ($b = 32.6$ \kms) 
for the neighbor (Fig.~\ref{fg4}).

\begin{figure}
\vspace{0.0cm}
\hspace{-2.2cm}\psfig{figure=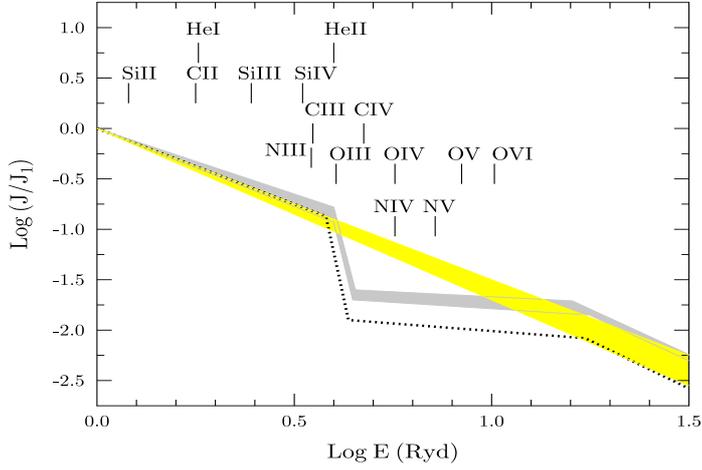,height=15cm,width=13.0cm}
\vspace{-4.0cm}
\caption[]{
SEDs complying with the line profiles from
the \zabs\ = 1.9410 system: yellow cone plots simple power laws, $J_\nu \propto \nu^{-\alpha}$,
with $\alpha$ = 1.5--1.7; grey shadowed area marks broken power laws with a break at 4 Ryd. 
The black dotted line 
shows an initial shape used in the SED adjustment procedure.
}
\label{fg5}
\end{figure}

Firstly, simple power laws, $J_\nu \propto \nu^{-\alpha}$, with 
the index $\alpha$ ranging from 0.6 to 1.8 were tried 
as a continuum shape in the EUV range. At $U$ determined by the measured ratio of $N$(\ion{C}{iv})/$N$(\ion{C}{ii}),
the column densities of silicon ions are reproduced with power-law continua having indices $\alpha = 1.5 - 1.7$, 
with the element abundances estimated as
[C] = -0.3 -- -0.4 and [Si/C] $\sim 0.0$ -- 0.05. 
However, these continua predict small fractions of \ion{O}{iii}, and 
one needs a rather strong overabundance of oxygen, [O/C] $\sim 0.5$, 
to describe the observed profiles of the \ion{O}{iii} lines. 
Such an overabundance is only marginally
consistent with measurements both in metal-rich stars (metal content $Z \ga 0.5\,Z_\odot$) and in high-metallicity 
\ion{H}{ii} regions that usually show [O/C] $< 0.3$
(Akerman \etal\ 2004; Garcia-Rojas \etal\ 2004; Bensby \& Feltzing 2006; Bresolin 2007; Esteban \etal\ 2009).
That is why more complex continuum shapes were tried as well: namely, broken power laws defined by three indices and
two breaking points (five factors) in the range 1~Ryd $< E <$ 15~Ryd (Fig.~\ref{fg5}). 
The boundary to the
soft X-ray range was fixed at 0.2 keV (14.7 Ryd) and $\alpha = 1.8$ at $E > 0.2$ keV 
(this last
setting is arbitrary since it does not affect the fractions of ions relevant for the present study).

At $U$ corresponding to the measured ratio of
$N$(\ion{C}{iv})/$N$(\ion{C}{ii}), the first-guess SED (dotted line in Fig.~\ref{fg5})  underproduced
the measured $N$(\ion{Si}{iv})/$N$(\ion{Si}{iii}) 
(corresponding \ion{Si}{iv} profiles are shown in the inserts in Fig.~\ref{fg3}).  
The latter was
chosen as a response function and the search in the five-factor space was performed for the direction toward
SEDs with a higher response. Moving along this direction in steps, we found the range of SEDs reproducing
both the observed $N$(\ion{C}{iv})/$N$(\ion{C}{ii}) and $N$(\ion{Si}{iv})/$N$(\ion{Si}{iii}). 
However, for some SEDs from this range 
we obtained a negative abundance ratio [Si/C] $< 0$.
Such an underabundance of silicon can occur
in the gas enriched mostly by the AGB stars (Garnett \etal\ 1995; Agafonova \etal\ 2011), 
but it is coupled both with a high (solar to oversolar) content of carbon and  
a relative overabundance of C to oxygen, [C/O] $> 0$ 
(Kingsburgh \& Barlow, 1999). 
In the present system, carbon abundance for any type of the ionizing spectrum does not 
exceed half-solar value. However, to describe the profiles of the
strong \ion{O}{iii} $\lambda\lambda832, 702, 507$ \AA\ lines,
the relative oxygen content should always be at least not lower than that of carbon,  i.e., 
[O/C] $> 0$.
Thus, we can exclude here AGB stars as a prevalent enrichment source and set zero as a lower limit
for the relative abundance of silicon to carbon: [Si/C] $>0$.
Using this condition as a new response function, 
we performed a second iteration of the continuum shape adjustment.

The range of acceptable SEDs finally obtained, 
i.e., those that allow us to describe self-consistently all observed
profiles along with reasonable abundance ratios, is shown by the shadowed area in Fig.~\ref{fg5}.
The physical parameters derived with SEDs from this range are listed in Table~\ref{fg2}.
The corresponding synthetic absorption-line
profiles are plotted by red in Figs.~\ref{fg3} and \ref{fg4}.
Compared to pure power laws, the SEDs with a break at $E \sim 4$ Ryd give a
fraction of \ion{O}{iii} that is almost twice as large, 
so that now the observed intensities of \ion{O}{iii} lines can be
well reproduced with a moderate oxygen-to-carbon overabundance ratio, 
[O/C] = 0.2--0.3. This behavior of the \ion{O}{iii}
fraction is easy to explain: the ionization potential of \ion{O}{iii} is 4.038 Ryd and a sharp decrease of the
number density of the ionizing photons just above this energy hampers 
the transition from \ion{O}{iii} to \ion{O}{iv}, thus retaining the oxygen in the double-ionized stage.

\subsection{Absorption complex at $z = 1.8916$ }
\label{sub-sect-4-2}

This system was described in detail by K\"ohler \etal\ (1999). Here we re-analyze it with
the aim of restoring the continuum shape of the UV ionizing radiation.

\begin{figure}
\vspace{0.0cm}
\hspace{-0.5cm}\psfig{figure=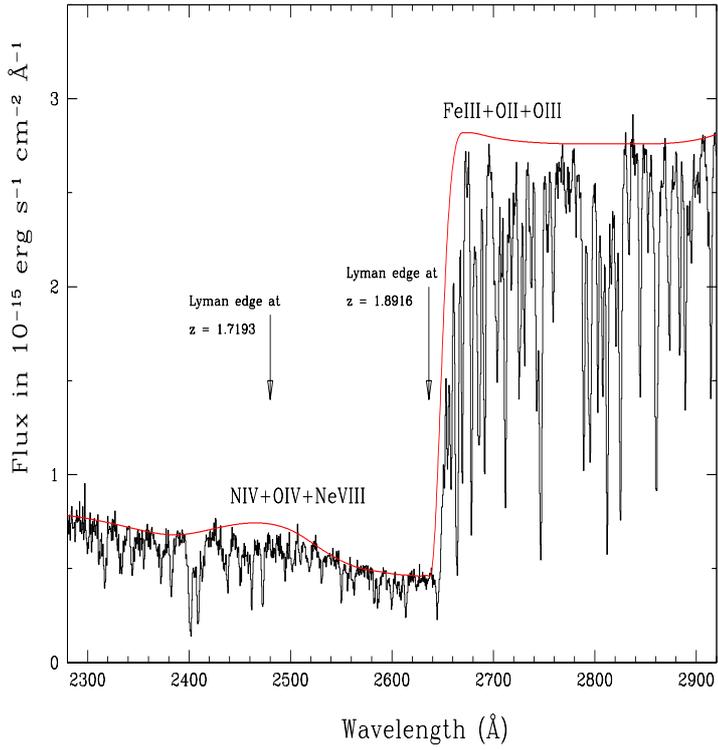,height=13cm,width=10.0cm}
\vspace{-2.0cm}
\caption[]{
Observed flux of \object{HS~1103+6416} obtained with the FOS and GHRS onboard the HST.
The arrows mark the \ion{H}{i} Lyman limits for the $z=1.8916$ and $z=1.7193$ systems.
}
\label{fg6}
\end{figure}

\begin{figure*}
\vspace{0.0cm}
\hspace{0.0cm}\psfig{figure=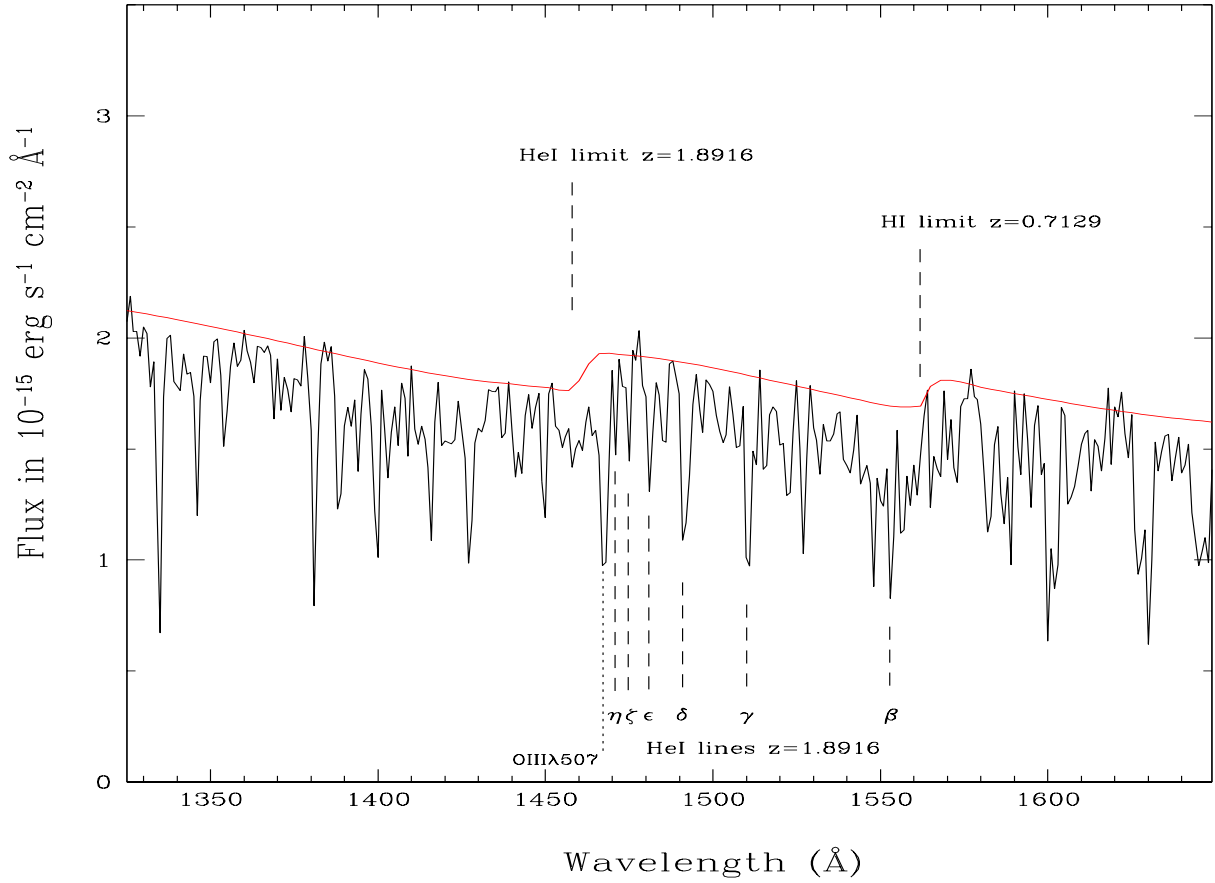,height=13cm,width=18cm}
\vspace{0.0cm}
\caption[]{
\ion{He}{i} Lyman limit at $z=1.8916$ and the identified \ion{He}{i} Lyman series lines.
The \ion{H}{i} Lyman limit at $z = 0.7129$ and the \ion{O}{iii} $\lambda507.388$ \AA\ line at
$z=1.8916$ are also indicated. The latter is blended with \ion{He}{i} Ly-$\theta$ ($\lambda507.718$ \AA)
and Ly-$\iota$ ($\lambda507.058$ \AA) lines.
}
\label{fg7}
\end{figure*}

The available UV parts of the \object{HS~1103+6416} spectrum 
contain clear Lyman limits for both \ion{H}{i} and \ion{He}{i}, 
making it possible to measure the corresponding column densities\footnote{The photoionization cross section 
for the \ion{He}{i} continuum is taken from Samson \etal\ (1994).}: 
$N$(\ion{H}{i}) $= (2.87 \pm 0.03)\times10^{17}$ \cm\ (Fig.~\ref{fg6}) 
and $N$(\ion{He}{i}) $= (1.5 \pm 0.3)\times10^{16}$ \cm\ (Fig.~\ref{fg7}).
We note that this is the first discovery of the \ion{He}{i}
continuum in any cosmic object except the Sun.
The spread of the $N$(\ion{H}{i}) values is determined by the uncertainty in the emission 
profile of the blend \ion{O}{ii}+\ion{O}{iii}+\ion{Fe}{iii} blueward of the \ion{H}{i} Lyman
edge and for $N$(\ion{He}{i}) by the noise in the spectrum. 
The ratio $N$(\ion{He}{i})/$N$(\ion{H}{i})  is sensitive to the radiation
intensity ratio at 1 Ryd and 1.8 Ryd (Lyman limit for \ion{He}{i}) 
and in principle could be used to estimate the slope of the continuum spectrum in this
energy range, similar to the way
$N$(\ion{He}{ii})/$N$(\ion{H}{i}) is used to probe the $J_1/J_4$ ratio in 
the \ion{H}{i}+\ion{He}{ii} Ly-$\alpha$ forest measurements.
But for a given ionization spectrum, the value of 
$N$(\ion{He}{i})/$N$(\ion{H}{i}) depends strongly on the ionization parameter, i.e., any
estimation of this ratio can be made only by involving metal lines.

\begin{figure*}
\vspace{0.0cm}
\hspace{-1.0cm}\psfig{figure=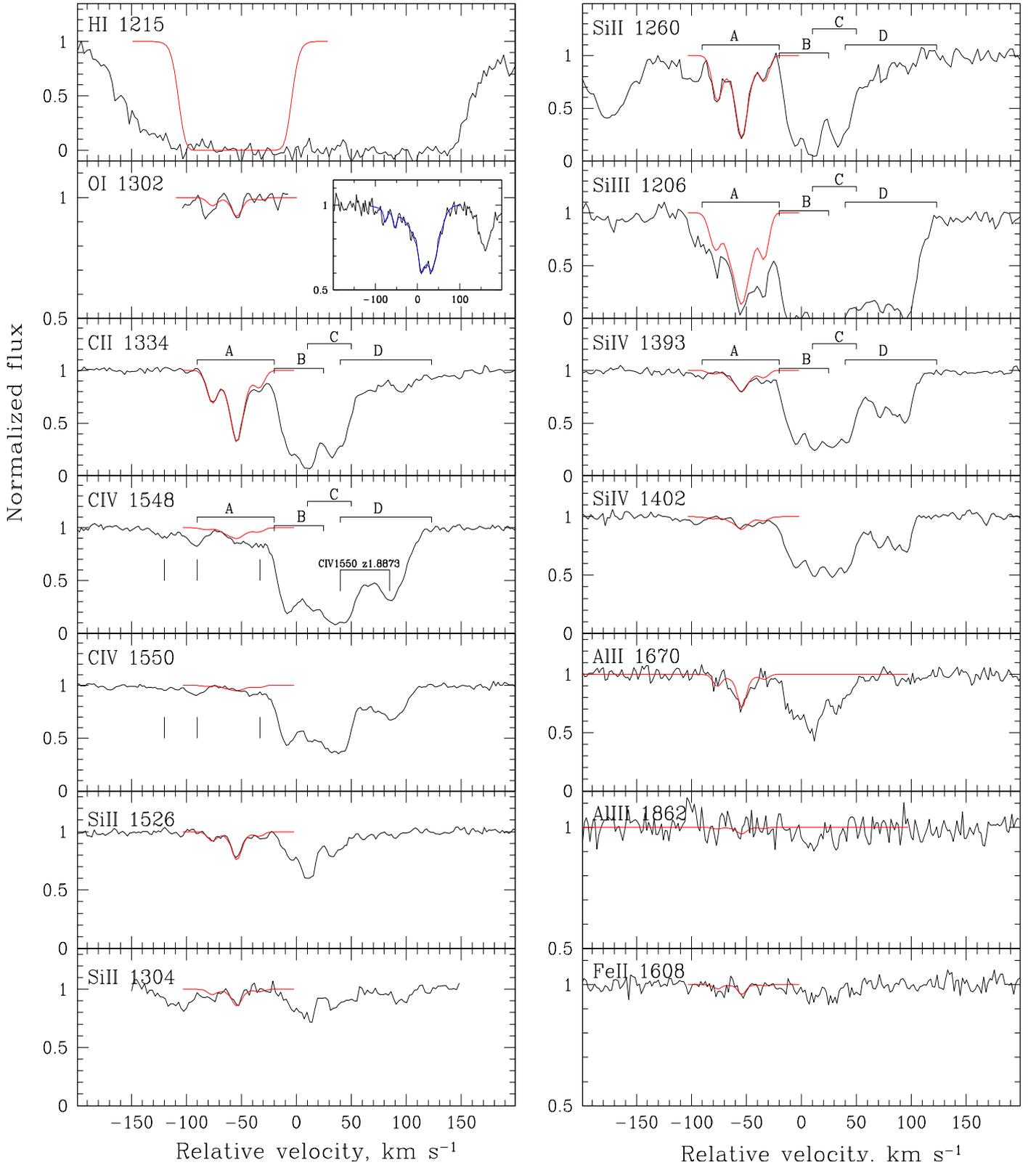,height=22cm,width=20cm}
\vspace{0.5cm}
\caption[]{
High-resolution lines observed in the $z = 1.8916$ system. 
Brackets indicate the sub-systems $A$, $B$, $C$, and $D$
with strong lines, vertical ticks in the \ion{C}{iv} panels point to superimposed sub-systems with 
weak absorption lines.
The synthetic profiles for metal ions (red) in the sub-system $A$ are calculated 
with broken power-law SEDs from the grey shadowed area in Fig.~\ref{fg10}.
The black line in the \ion{O}{i} panel is a restored profile from the forest absorption shown in the insert.
The black and blue curves in the insert are the original and deconvolved profiles, respectively.
}
\label{fg8}
\end{figure*}

The absorption lines of metal ions detected in the Keck spectrum are shown in Fig.~\ref{fg8}. 
Essential for the SED reconstruction are weak but clearly present lines of
\ion{Fe}{ii} $\lambda 1608$ \AA\ (other \ion{Fe}{ii} transitions are beyond the observed range) and
\ion{O}{i} $\lambda 1302$ \AA. The latter is blended with an \ion{H}{i} line from the Ly-$\alpha$
forest, but the component that falls into the blue wing of \ion{H}{i} can be deconvolved. 
Although the complex represents a superposition of
several individual absorbers (marked by the letters $A, B, C$, and $D$ in Fig.~\ref{fg8}), 
they can be  separated if we compare the relative strengths 
of their \ion{C}{ii}, \ion{C}{iv},
\ion{Si}{ii}, \ion{Si}{iii}, and \ion{Si}{iv} lines:
strong \ion{C}{ii}, \ion{Si}{ii} along with weak 
\ion{C}{iv}, \ion{Si}{iv} lines characterize the gas in the absorber $A$; 
comparably strong pairs \ion{C}{ii}, \ion{C}{iv}, and 
\ion{Si}{ii}, \ion{Si}{iv} are revealed in the absorber $B$; and 
lines of highly ionized species
\ion{C}{iv} and \ion{Si}{iv} are slightly stronger than low-ionization 
\ion{C}{ii}, \ion{Si}{ii} in the absorbers $C$ and $D$.
A significantly different metal content in the sub-systems $C$ and $D$ is
conceivable. Additionally, there are superimposed absorbers seen mostly in weak 
\ion{C}{iv} (in some cases also in \ion{Si}{iv}) transitions (indicated by vertical ticks in Fig.~\ref{fg8}).

In order to distribute between the individual absorbers the total column density 
$N$(\ion{H}{i}) measured from the optical depth at the Lyman edge, we fit all available
\ion{H}{i} lines to the sum of Voigt components, each of which is described by
three parameters: the line center, the broadening $b$-parameter, and the column density. 
However, unlike the previous system at $z = 1.9410$, which
consists of two well-separated absorbers with the total 
$N$(\ion{H}{i}) $\sim 5\times10^{15}$ \cm, we have here at least four close and partially
overlapping absorbers 
with the total $N$(\ion{H}{i}) $\sim 3\times10^{17}$ \cm. 
Coupled with a low resolution (FWHM $\sim 200$ \kms)
of higher Lyman series lines of \ion{H}{i}, this
makes the deconvolution procedure
numerically unstable (ill-posed), which results in multiple solutions 
(i.e., in significantly different column densities for the same component, depending 
on model, initial guesses, exit conditions,  etc.), even if the total 
$N$(\ion{H}{i}) remains constant.

\begin{figure*}
\vspace{0.0cm}
\hspace{-0.7cm}\psfig{figure=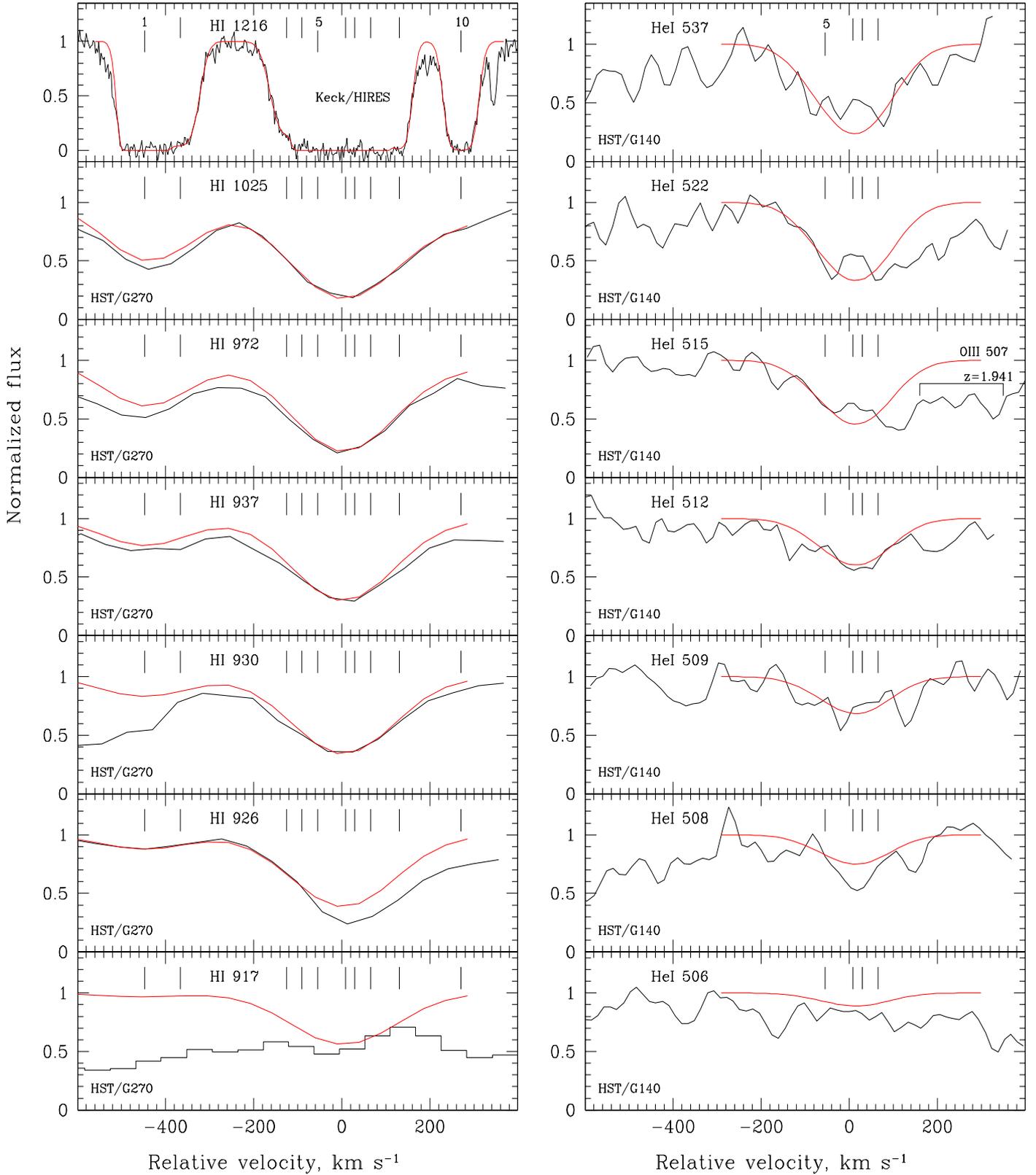,height=22cm,width=20cm}
\vspace{0.0cm}
\caption[]{
Low-resolution \ion{H}{i}  and \ion{He}{i} lines from the $z = 1.8916$ system. 
Tick marks indicate the positions of individual sub-components
in the \ion{H}{i} profile deconvolution. The synthetic profiles
are shown by red and the observed ones by black curves. 

}
\label{fg9}
\end{figure*}

A common approach to regularizing an ill-posed problem is to place some a priori known 
constraints on the fitting parameters. In the present
case such constraints can be derived from metal lines: the absorber $B$ has 
$N$(\ion{Si}{ii}) = $2.2\times10^{13}$ \cm\ and 
$N$(\ion{Si}{iv}) = $2.1\times10^{13}$ \cm\ 
(Table~2 in K\"ohler \etal\ 1999).
The  \ion{Si}{iii} $\lambda1206$ \AA\ line is
saturated (Fig.~\ref{fg8}) and hence should have
$N$(\ion{Si}{iii}) $\sim 10^{14}$ \cm. The line is in the Ly-$\alpha$ forest and 
in principle could be blended with some forest feature.
However, any strong Ly-$\alpha$ blend with $N$(\ion{H}{i}) $\sim$ $10^{14}$ \cm\
is excluded by the continuum window 
at the expected position of Ly-$\beta$ $\lambda 1025$ \AA, and there are
no candidate systems that could produce strong metal lines at the position of \ion{Si}{iii}. 
Thus the observed saturation is indeed
due to absorption in \ion{Si}{iii} $\lambda1206$ \AA, although contamination with a weak, 
$N$(\ion{H}{i}) $\sim$ $10^{13}$ \cm, hydrogen line (or lines) cannot be excluded (see the 
synthetic \ion{Si}{iii} profile in Fig.~\ref{fg8}).
The silicon ion fractions depend quite
strongly on the gas metallicity $Z$, and trial calculations with CLOUDY have shown that 
for a wide range of shapes of the ionizing continua
the required large fraction of \ion{Si}{iii} at
\ion{Si}{ii}/\ion{Si}{iv} $\sim 1$ can be realized only if $[Z] \la -0.4$. 
Similarly, in the absorber $D$ the column density of the double-ionized silicon 
$N$(\ion{Si}{iii}) $\sim (3-4)\times10^{13}$ \cm\ at $N$(\ion{Si}{iv})/$N$(\ion{Si}{ii}) $\ga 3$ 
is reproduced only at low gas metalicities, $[Z] \la -1.0$. 
It follows from these conditions that the sub-systems $B$ and $D$ should have 
$N$(\ion{H}{i}) $\sim 10^{17}$ \cm. Compared to the absorber $B$, the absorber $C$ reveals two times lower
column densities of \ion{Si}{ii} and \ion{C}{ii}, 
but almost similar $N$(\ion{Si}{iv}) and $N$(\ion{C}{iv})\ (K\"ohler \etal\ 1999). 
Assuming for $C$ 
the same metallicity as in $B$ (which looks plausible taking into account the closeness 
and similarity of line profiles), 
we estimate for it $N$(\ion{H}{i})  $\sim 5\times10^{16}$ \cm. 
These constraints on the \ion{H}{i} column densities, along with setting the bounds for
the component centers close to the position of the strongest metal absorption, 
allow us to stabilize the \ion{H}{i} deconvolution
and to obtain the parameters listed in Table~\ref{tbl-3}. 
The synthetic \ion{H}{i} profiles calculated with these parameters
are shown by red curves in Fig.~\ref{fg9}.

The sub-system $A$ is optically thin both in \ion{H}{i}, 
$N$(\ion{H}{i}) $\sim 4\times10^{16}$ \cm\ and \ion{He}{ii}, 
$N$(\ion{He}{ii})  $\sim 2\times10^{17}$ \cm;  with its
great diversity of ionic transitions, this sub-system is suitable for the SED reconstruction. 
The measured column densities are listed in Table~\ref{tbl-1}. 
The red wings of several lines in the sub-system $A$ are blended with corresponding components 
from the other sub-systems (Fig.~\ref{fg8}),
but \ion{Si}{ii} and \ion{Si}{iv} are affected only weakly and can be used to estimate
the ionization parameter.   
The power-law SEDs with $\alpha = 0.9 - 1.2$ reproduce the column densities of all ions 
with the following abundances: [C] = -0.45 -- -0.55, [Al,Si,Fe/C] = 0.0 -- 0.1, [O/C] = 0.0 -- 0.3,
which are consistent with standard models of chemical enrichment. 
The profiles of both \ion{C}{iv} $\lambda 1548$ \AA\ and \ion{C}{iv} $\lambda 1550$ \AA\ lines
come out slightly underestimated, but they might be blended with components from
weak-\ion{C}{iv} sub-systems like those seen at $v = -120$ or $-90$  \kms\ (tick marks
in the \ion{C}{iv} panels in Fig.~\ref{fg8}). 
However, these power laws also significantly underproduce 
\ion{C}{iv} profiles in the adjacent system $B$, which are
much stronger and, hence, less affected by possible blends. 
Thus pure power-law continua are probably not the best choice for the present absorber.

Softer power laws ($\alpha \sim 1.5$), like those considered for the $z = 1.9410$ system, 
reproduce all ions including \ion{C}{iv}, but only at the expense of 
high overabundances of O and Fe relative to C: [C]= -0.6 -- -0.7, [Si/C] = 0.15 -- 0.20, [O,Fe/C] $> 0.5$. 
A simultaneous overabundance of both iron and $\alpha$-elements (O, Si) 
to carbon at a quite high carbon content is neither predicted nor observed. 
Besides, the total column densities of \ion{C}{ii} and \ion{O}{ii}, 
$N$(\ion{C}{ii}) $= 3.3\times10^{14}$ \cm\ 
and $N$(\ion{O}{ii}) $= (8.5-9.0)\times10^{14}$ \cm\  (see Fig.~\ref{fg2}), 
do not indicate a high oxygen overabundance. Thus, we can conclude
that pure power-law SEDs,
if adopted as the ionizing continua, are significantly different at $z=1.8916$ and 1.9410. 

The continuum shape from the previous $z = 1.9410$ system 
(upper boundary of the shadowed region in Fig.~\ref{fg5}) was tried
as an initial guess for a broken power-law SED. 
With this SED, the \ion{Fe}{ii} $\lambda 1608$ \AA\ and \ion{O}{i} $\lambda 1302$ \AA\ lines
can also be reproduced with [O,Fe/C] $> 0.5$. 
In general, ion fractions of \ion{Fe}{ii} and \ion{O}{i} are only weakly
influenced by the continuum shape at $E > 1$ Ryd and depend mostly on the ionization parameter $U$, 
which starts to decrease sharply at $\lg U \ga -4$.
Thus, to obtain reasonable relative abundances of Fe and O, we have to modify the initial continuum shape so
that the measured ratio $N$(\ion{Si}{ii})/$N$(\ion{Si}{iv}) (along with a sufficient amount of \ion{C}{iv})
will be achieved at a possible low $U$. 

\begin{figure}
\vspace{0.0cm}
\hspace{-2.2cm}\psfig{figure=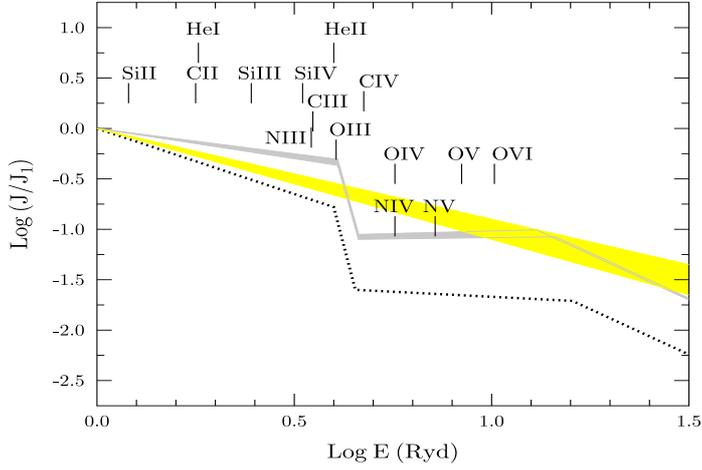,height=15cm,width=13.0cm}
\vspace{-4.0cm}
\caption[]{
SEDs complying with the lines observed 
in the $z = 1.8916$ system: yellow cone plots simple power laws, $J_\nu \propto \nu^{-\alpha}$,
with $\alpha$ = 0.9--1.2; grey shadowed area indicates broken power laws with a break at 4 Ryd. 
The black dotted line 
shows an initial shape used in the SED adjustment procedure; here it is chosen as an upper
bound of the broken power-law SEDs restored for the $z = 1.9410$ system (grey shadowed area in Fig.~\ref{fg5}).
}
\label{fg10}
\end{figure}

The range of the broken power-law SEDs acceptable for the present system is shown by
the shadowed area in Fig.~\ref{fg10}, with the synthetic line profiles plotted in red in Fig.~\ref{fg8} 
and the estimated abundances given in Table~\ref{tbl-2}. 
These continua are very hard, with a slope between 1 Ryd and 4 Ryd corresponding 
to $\alpha = 0.5 - 0.6$ (cf. with $\alpha = 1.3 - 1.5$ at $z = 1.9410$). 
We note that K\"ohler \etal\ (1999),
in spite of their quite different approach, also proposed a continuum shape with $\alpha = 0.5$
and an order of magnitude break in the intensity at 4 Ryd as a favorable SED for 
the present system. 

Such hard slopes at $E > 1$ Ryd are indeed observed in some AGNs (Scott \etal\ 2004; Shull \etal\ 2012). 
This type of objects includes a well-known quasar
\object{HE~2347--4342} (toward which the \ion{He}{ii} Ly-$\alpha$ forest was firstly resolved); 
according to Telfer \etal\ (2002) it should have $\alpha$ = 0.56.       

Now, using the \ion{H}{i} column densities for the individual components (Table~\ref{tbl-3}), 
the restored SED, and the column densities for metal
ions from K\"ohler \etal\ (1999) to estimate the ionization parameters in the sub-systems $A$, $B$, $C$ and $D$, 
we can calculate the total amount of \ion{He}{i} and compare it with the value given by the \ion{He}{i} Lyman limit
optical depth.
Assuming the helium abundance (in number) of 0.083, we obtain 
$N$(\ion{He}{i})$_A = (2.5 - 3.0)\times10^{15}$ \cm, 
$N$(\ion{He}{i})$_B = (5.0 - 6.0)\times10^{15}$ \cm, 
$N$(\ion{He}{i})$_C = (2.0 - 2.5)\times10^{15}$ \cm, and
$N$(\ion{He}{i})$_D = (2.5 - 3.0)\times10^{15}$ \cm, which results 
in the total 
$N$(\ion{He}{i})$_{\rm tot} = (1.2 - 1.5)\times10^{16}$ \cm\ complying well
with the value estimated from the \ion{He}{i} Lyman edge. 

The G140 part of the \object{HS~1103+6416}
spectrum also contains a series of \ion{He}{i} Lyman lines. 
To model these lines with a sum of components,
we need a broadening parameter for each component.  
For the component $A$, we obtain 
$b_{\scriptstyle \rm He\,\scriptscriptstyle I} = 15$ \kms\ from the
comparison of the \ion{He}{i} synthetic profile 
calculated with the density-velocity distributions estimated from metal lines 
and a single Voigt profile with  the same $N$(\ion{He}{i}). This means that
the broadening is dominated by turbulent effects. Assuming the same ratio ($\sim 0.9$) between
the $b$-parameters of the \ion{H}{i} and \ion{He}{i} lines in the
components $B, C$, and $D$, we model the \ion{He}{i} profiles 
as the sum of four Gaussians convolved with the instrumental function. 
The synthetic \ion{He}{i} profiles are shown by red curves in the right-hand panels in Fig.~\ref{fg9}.

\subsection{Absorption system at $z = 1.8873$ }
\label{sub-sect-4-3}

The system is detached from the $z = 1.8916$ absorber by only 450 \kms, which corresponds to $\sim 2.2$ Mpc
if the offset is purely cosmological.

\begin{figure*}
\vspace{0.0cm}
\hspace{0.0cm}\psfig{figure=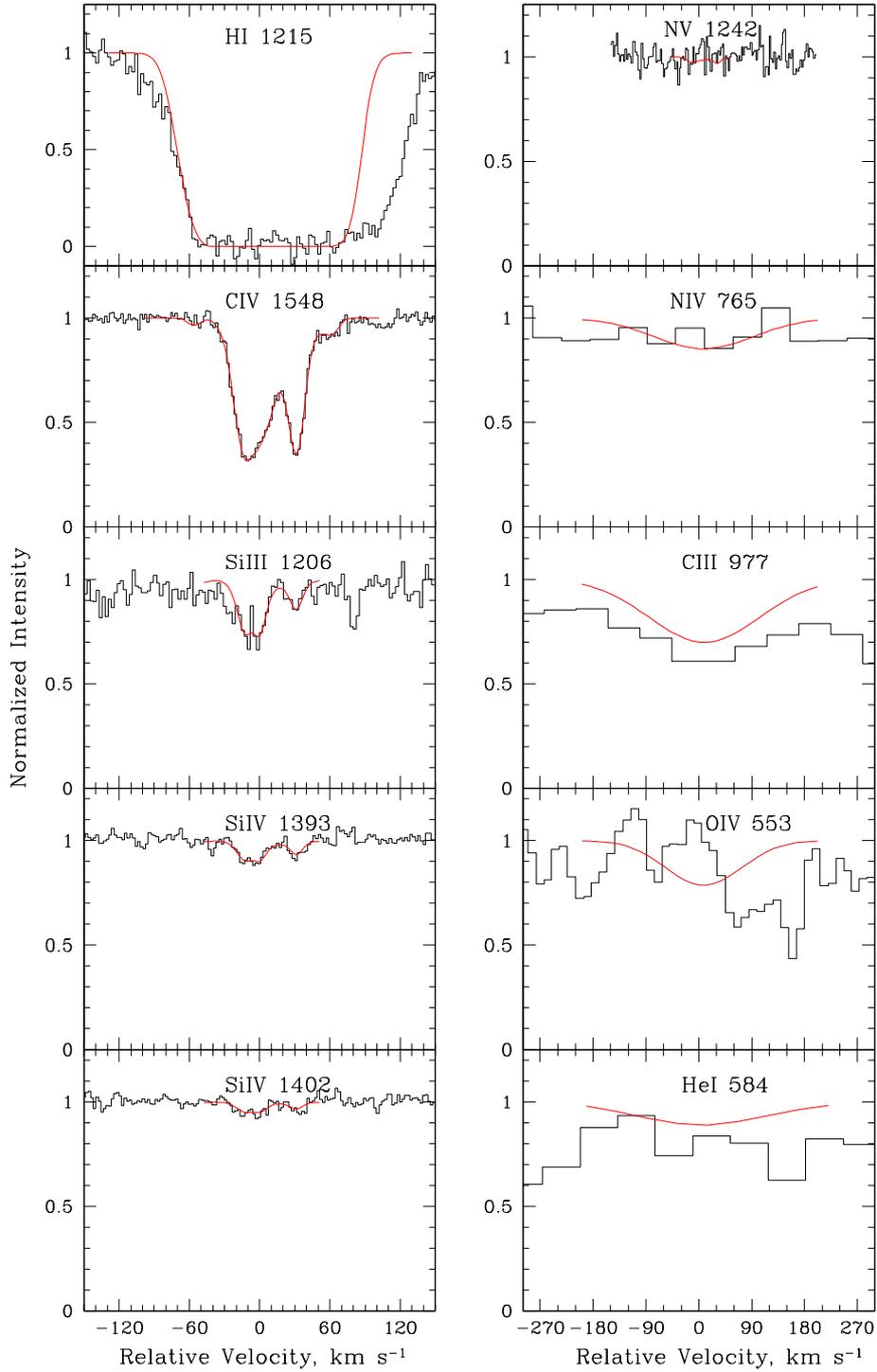,height=20cm,width=20.0cm}
\vspace{0.0cm}
\caption[]{
Absorption lines observed in the $z = 1.8873$ system.
The synthetic profiles (red) are calculated  with the 
SEDs shown  in Fig.~\ref{fg12}.
}
\label{fg11}
\end{figure*}

The metal ions are represented by lines of 
\ion{C}{iv} $\lambda1548$ \AA\ (\ion{C}{iv} $\lambda1550$ \AA\ is blended with 
\ion{C}{iv} $\lambda1548$ \AA\ from the $z=1.8916$ system ) and
\ion{Si}{iii} $\lambda1206$ \AA, \ion{Si}{iv} $\lambda\lambda1393, 1402$ \AA\ (Fig.~\ref{fg11}). 
There is a weak absorption at the position of \ion{C}{ii} $\lambda1334$ \AA, 
but it is not clear whether it is indeed \ion{C}{ii} or some forest feature since 
the spectrum of \object{HS~1103+6416} contains 
many weak and unidentified lines in the vicinity of this wavelength.
In the UV part, a continuum window seen at the position of the \ion{O}{iv} $\lambda553$ \AA\ line
and weak absorption at the position of \ion{N}{iv} $\lambda765$ \AA\ 
can be used to set upper limits on their column densities.
The hydrogen content was estimated in the previous section and is given in 
Table~\ref{tbl-3} (component No.~1);
the column densities for metal ions are listed in Table~\ref{tbl-1}.

Power-law SEDs with $\alpha > 1.6$ reproduce the measured ion ratios with 
[C] $\sim -1.0$ and relative underabundance of silicon, [Si/C] $< 0$, and thus can be ruled out
for the same reasons as in Sect.~\ref{sub-sect-4-1}. 
Power laws with $\alpha = 1.6 - 1.8$
give [C] $\sim -1.2$ and [Si/C] $\sim 0 - 0.2$, i.e., C and Si abundances are in line with
standard chemical enrichment. However, under the assumption of [O/C] $\sim 0.2 - 0.3$, which is also
in line with standard enrichment, these SEDs predict quite a large column density of 
$N$(\ion{O}{iv}) = $(5.0 - 6.0)\times10^{14}$ \cm,
which is only marginally consistent with the 
noisy continuum at the expected position of \ion{O}{iv} $\lambda553$ \AA.

Broken power-law SEDs reconstructed for the previous $z=1.8916$ system are definitely
inconsistent with lines in the present absorber: at $U$ estimated from the condition
$N$(\ion{Si}{iv}) = $N$(\ion{Si}{iii}),
the predicted intensity of the \ion{C}{ii} $\lambda 1334$ \AA\ line significantly overestimates the
observed profile; besides, the measured $N$(\ion{C}{iv}) requires the abundance ratio [Si/C] $\sim -0.5$. 
Thus, we can conclude that the SED at $z=1.8873$ in any case {\it differs} from that at $z=1.8916$.

Due to a small number of available lines at $z = 1.8873$, it is not possible to restrict the range of
appropriate continuum shapes as was done in Sects.~\ref{sub-sect-4-1} and \ref{sub-sect-4-2}. 
A large plurality of SEDs
consistent both with the measured column densities and with standard chemical enrichment
(i.e., delivering [O,Si/C] $> 0$ at [C] $\sim -1.0$) can be proposed; two examples are shown
in Fig.~\ref{fg12} by solid lines. It is seen that the SEDs at $z=1.8873$ are considerable softer,
either only at $E > 4$ Ryd or in the whole EUV range, than the ionizing continuum at $z=1.8916$.

\subsection{Absorption system at $z = 1.7193$ }
\label{sub-sect-4-4}

\begin{figure}
\vspace{0.0cm}
\hspace{-2.2cm}\psfig{figure=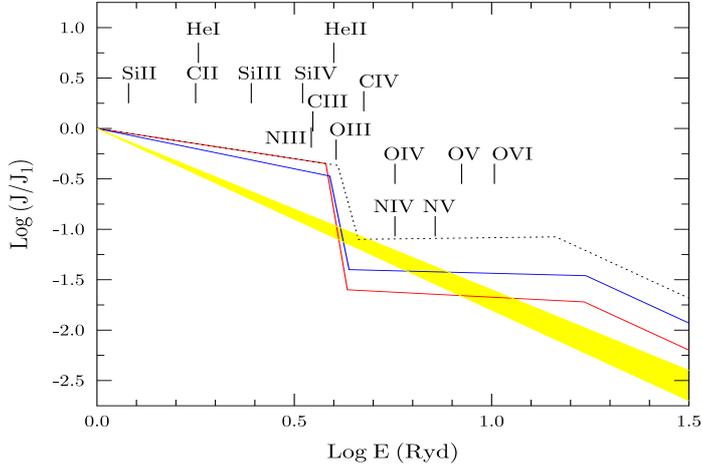,height=15cm,width=13.0cm}
\vspace{-4.0cm}
\caption[]{
SEDs restored for the $z = 1.8873$ system: yellow cone plots simple power laws, $J_\nu \propto \nu^{-\alpha}$,
with $\alpha$ = 1.6--1.8; red and blue lines show two examples of broken power laws. The black dotted line indicates
the SED restored for the previous $z = 1.8916$ system; this SED does not comply with the
lines observed at $z = 1.8873$.
}
\label{fg12}
\end{figure}

\begin{figure*}
\vspace{0.0cm}
\hspace{-1.0cm}\psfig{figure=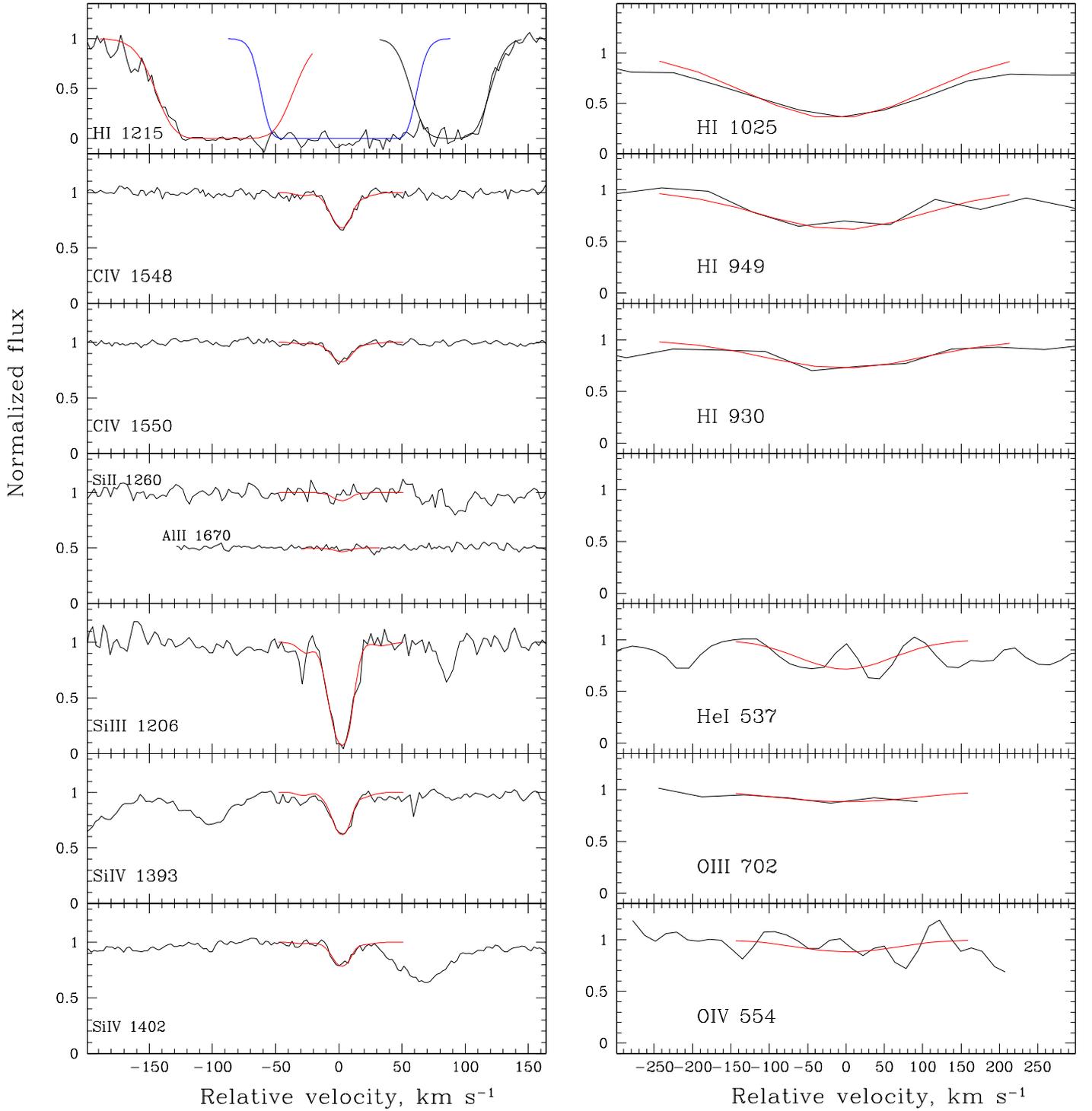,height=20cm,width=20.0cm}
\vspace{0.0cm}
\caption[]{
Absorption lines observed in the $z = 1.7193$ system.
The synthetic profiles (red) are calculated  with SEDs from the 
shadowed area in Fig.~\ref{fg15}.
}
\label{fg13}
\end{figure*}

The system is detached by 17400 \kms, or $\sim 90$ Mpc, from the previous 
$z=1.8873$ absorber. In the Keck spectrum, the \ion{H}{i} $\lambda1215$ \AA,
\ion{C}{iv} $\lambda\lambda1548, 1550$ \AA, 
\ion{Si}{iii} $\lambda1206$ \AA, \ion{Si}{iv} $\lambda\lambda1393, 1402$ \AA\ lines are present, 
with continuum windows at the expected positions of \ion{Si}{ii} $\lambda1260$ \AA,
\ion{Al}{ii} $\lambda1670$ \AA, and \ion{N}{v} $\lambda1242$ \AA\ (Fig.~\ref{fg13}).  
The \ion{C}{ii} $\lambda1334$ \AA\ line is blended with a strong forest line, and it is unclear 
whether there is a corresponding absorption or not.  
The UV spectrum contains a series of \ion{H}{i}  Lyman lines up to the Lyman limit at 2479 \AA\ 
(marked by the arrow in Fig.~\ref{fg6}). 
Fitting the \ion{H}{i} profiles to the three component model, 
we obtain $N$(\ion{H}{i}) $= (1.5-1.6)\times10^{16}$ \cm\ 
for the total hydrogen 
and $N$(\ion{H}{i}) $= (1.4-1.5)\times10^{16}$ \cm\ for the central component 
exhibiting metal lines.
The column densities measured from the \ion{Si}{iii} $\lambda1206$ \AA, 
\ion{Si}{iv} $\lambda\lambda1393, 1402$ \AA, and \ion{C}{iv} $\lambda\lambda1548, 1550$ \AA\ profiles 
are given in Table~\ref{tbl-1}.
A highly unusual ratio \ion{Si}{iv}/\ion{C}{iv} $\sim 0.5$ at  
$\log U \sim -2$ (determined by \ion{Si}{iii}/\ion{Si}{iv} $\sim 2$
and \ion{Si}{iv}/\ion{Si}{ii} $> 13$) is found: in the systems at 
$z=1.9410$ and $z=1.8873$ with comparable ionization states of the absorbing
gas, the \ion{Si}{iv}/\ion{C}{iv} ratio is only $\sim 0.06$ and 0.03, respectively.
The measured ratios
\ion{Si}{ii}/\ion{Si}{iv}, \ion{Si}{iii}/\ion{Si}{iv}, and \ion{Si}{iv}/\ion{C}{iv} 
are neither reproduced by
simple power-law SEDs ($\alpha = 0.5 - 1.9$) 
nor by any of the broken power laws restored from the previous absorption systems. Besides,
the element abundances compliant with the \ion{Si}{iv}/\ion{C}{iv} ratio require
[Si/C] $> 0.6$ at [C] $< -2.0$.
However, at such a low [C], the overabundance
of Si to C is well constrained by [Si/C] $\la 0.3$ (Erni \etal\ 2006; Cooke \etal\ 2011).
The search for direction (in five-factor space)
toward SEDs ensuring both the measured ionic ratios and [Si/C] $\la 0.3$ was performed
with the same initial guess as for the $z=1.9410$ system (Fig.~\ref{fg5}) and with the response function defined
as a [Si/C] value (first iteration) and a
\ion{Si}{iv}/\ion{Si}{ii} ratio (second iteration). 
The resulting range is shown by the shadowed area  in Fig.~\ref{fg12};
the upper envelope gives [Si/C]  $\sim 0.3$ and the lower one is determined from the condition [Si/C] $\sim 0$.
The physical parameters obtained with SEDs from the shadowed area are listed 
in Table~\ref{tbl-2} and the synthetic spectra are shown
by red curves in Fig.~\ref{fg13}.

\begin{figure}
\vspace{0.0cm}
\hspace{-1.0cm}\psfig{figure=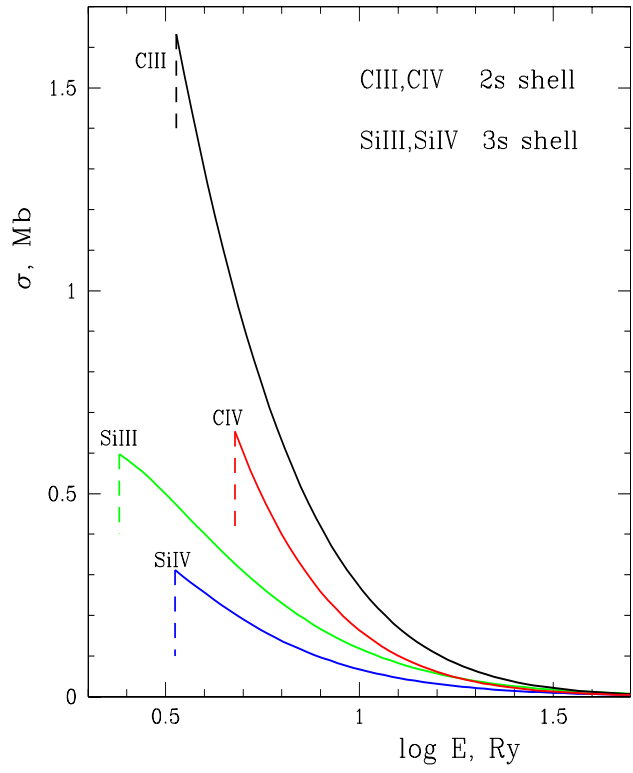,height=15cm,width=13cm}
\vspace{-3.0cm}
\caption[]{
Photoionization cross sections for the \ion{Si}{iii}, \ion{Si}{iv}, 
\ion{C}{iii}, and \ion{C}{iv} ions used in CLOUDY.
}
\label{fg14}
\end{figure}

The restored SEDs differ considerably from those derived from the previous absorption systems  
as they are much softer and, most striking, have a sharp break shifted to lower energies. 
Since this shift is a decisive
argument in the further interpretation of SEDs, it is important to ascertain that it is not a computational
artefact. 
The relative Si abundance is calculated as
[Si/C] $= \log$(\ion{Si}{iv}/\ion{C}{iv}) + 
$\log(\Upsilon_{\scriptstyle \rm C\,\scriptscriptstyle IV}/
\Upsilon_{\scriptstyle \rm Si\,\scriptscriptstyle IV})$ 
$- \log$(Si/C)$_\odot$, 
i.e., to ensure [Si/C]  $\sim 0.3$ the ionizing continuum
should provide at $U$ determined by 
\ion{Si}{iii}/\ion{Si}{iv} $\sim 2$ 
at least two times lower ratio 
$\log(\Upsilon_{\scriptstyle \rm C\,\scriptscriptstyle IV}/
\Upsilon_{\scriptstyle \rm Si\,\scriptscriptstyle IV})$ 
as compared to the initially tried SEDs.  
It is clear that to increase
the fraction of \ion{Si}{iv} we have to simultaneously push the ionization of 
\ion{Si}{iii} (i.e., increase the number
of photons with $E > 2.46$ Ryd, the ionization potential of \ion{Si}{iii}) 
and suppress the ionization of \ion{Si}{iv} (decrease the number of 
photons with $E > 3.32$ Ryd, the ionization potential of \ion{Si}{iv}). 
At the same time, the ionization of \ion{C}{iii} (the ionization potential 3.52 Ryd)
should also be suppressed to decrease the fraction of \ion{C}{IV}.
Figure~\ref{fg14} shows photoionization cross sections for 
\ion{C}{iii}, \ion{C}{iv}, \ion{Si}{iii}, and \ion{Si}{iv} as they are used in CLOUDY 
to calculate the ion fractions ($\Upsilon_i$): 
\ion{C}{iii} has the highest cross section with a
strong gradient, whereas the cross sections of \ion{Si}{iii} and 
\ion{Si}{iv} are smaller and more mildly sloping. 
This explains why the fractions of \ion{C}{iii} and \ion{C}{iv} are extremely 
sensitive to the distribution of energy in the range 3.5-4.0 Ryd,  
whereas the fractions of \ion{Si}{iii} and \ion{Si}{iv}
demonstrate a significantly weaker dependence. 
As a result of such behavior, the ratio 
$\Upsilon_{\scriptstyle \rm C\,\scriptscriptstyle IV}/
\Upsilon_{\scriptstyle \rm Si\,\scriptscriptstyle IV}$ 
can be lowered by two times only if the ionizing energy is strongly suppressed 
starting at the ionization threshold of \ion{C}{iii}, i.e., at $E \sim$ 3.5 Ryd.

In this regard we note that the modeling of metagalactic ionizing continua produced by the quasar
radiation transferred through the IGM  predicts at redshifts $z < 2$ that the \ion{He}{ii} 
breaks also shifted to lower energies (to the same 3.5 Ryd) due to the Hubble expansion 
(Haard \& Madau 1996, Fardal \etal\ 1998).
However, these breaks occur smeared out, and the corresponding SEDs cannot ensure
$\Upsilon_{\scriptstyle \rm C\,\scriptscriptstyle IV}/
\Upsilon_{\scriptstyle \rm Si\,\scriptscriptstyle IV}$ low enough to obtain [Si/C] $\sim 0.3$
for the measured \ion{Si}{iv} and \ion{C}{iv} column densities. 
For example, the ionizing continuum of Haard \& Madau (1996)
shown in Fig.~\ref{fg12} gives [Si/C]  $\sim 0.55$.

The absorber at $z=1.7193$ is optically thin in the \ion{H}{i} Lyman
continuum but opaque in \ion{He}{ii}. Due to the softness of the ionizing continuum, the
estimated $N$(\ion{He}{ii}) ranges from $2.5\times10^{18}$ \cm\ (upper envelope) 
to $6.0\times10^{18}$ \cm\ (lower envelope). 
This means that the incident radiation is attenuated within the absorber and is probably slightly
harder at $E > 4$ Ryd than the one restored from the absorption lines. 
However, local processes in the absorber itself do
not affect the energy range below 4 Ryd and thus cannot be responsible for the intensity 
suppression at 3.5 Ryd.

\subsection{Absorption system at $z = 1.1923$ }
\label{sub-sect-4-5}

Moving from the $z = 1.7193$ system down the redshift scale, the next system with metal lines is found
only at $z = 1.1923$, i.e., it is detached by $\sim 900$ Mpc (or $\sim 0.3$c in the velocity space). 
Some lines from this system were used to probe the wavelength calibration of the
\object{HS~1103+6416} spectrum and are shown in Fig.~\ref{fg2}. 
The presence of several strong Lyman series lines (\ion{H}{i} $\lambda\lambda1025, 972$ \AA) 
and a very shallow Lyman limit allow us to estimate
the hydrogen content as $N$(\ion{H}{i}) $\sim 5\times10^{15}$ \cm.
Except for the \ion{C}{iv} doublet, all other expected
metal lines, \ion{C}{iii} $\lambda977$ \AA, 
\ion{O}{iii} $\lambda\lambda832,702$ \AA, \ion{O}{iv} $\lambda787$ \AA,
\ion{Si}{iii} $\lambda1206$ \AA, and \ion{Si}{iv} $\lambda1393$ \AA, fall in the low-resolution
part of the spectrum. 
Their column densities can be estimated using the radial velocity structure found from the analysis of
the \ion{C}{iv} lines (see Sect.~\ref{sect-3}) and, hence, they are not very accurate. Besides, 
at FWHM $\sim 200$ \kms, it is always possible that some intervening absorption 
lines also contributed to the observed profile.
However, with $N$(\ion{C}{iv}) $= 2.2\times10^{14}$ \cm, we can expect the same order
of magnitude column densities for 
\ion{C}{iii}, \ion{O}{iii}, and \ion{O}{iv}, so that their possible blending with a weak Ly-$\alpha$ with
$N$(\ion{H}{i}) $\sim 10^{13}$ \cm\
(stronger hydrogen
lines are excluded by continuum windows at the corresponding positions of Ly-$\beta$) will not
significantly affect the measured column densities. 
Unlikely \ion{C}{iii}, \ion{O}{iii}, and \ion{O}{iv}, lines of silicon ions with
expected column densities of $\sim 10^{13}$ \cm\ fall in the absorption troughs and
their column densities cannot be even roughly estimated. 
The measured column densities of $N$(\ion{O}{iii}) $\sim (4.0 - 5.0)\times10^{14}$ \cm,
$N$(\ion{C}{iii}) $\sim (1.5 - 2.0)\times10^{14}$ \cm, and
$N$(\ion{O}{iv}) $\sim (1.0 - 1.5)\times10^{15}$ \cm\ 
are sufficient to rule out both the power laws
and soft SEDs of a type recovered at $z = 1.7193$ (i.e., with a sharp break shifted to E $\le 3.5$ Ryd) as
an ionizing continuum for the present system.  At $U$ 
determined by the ratio $N$(\ion{C}{iv})/$N$(\ion{C}{iii}) $\ge 1.0$,  
all these SEDs produce a small fraction of \ion{O}{iii} and, as a result, a large ratio
$N$(\ion{O}{iv})/$N$(\ion{O}{iii}) $\ge 10$, which significantly exceeds 
the observed $N$(\ion{O}{iv})/$N$(\ion{O}{iii}) $\sim 3$. 
The ionizing spectra with a break
at 4 Ryd similar to those shown in Figs.~\ref{fg5}, \ref{fg10}, and \ref{fg12} seem to be more appropriate
because they provide $N$(\ion{O}{iv})/$N$(\ion{O}{iii}) $\sim 3 - 5$.

\section{Discussion}
\label{sect-5}

Using lines of diverse metal ions, we reconstruct  
the SED of the environmental radiation field that ionizes
absorption systems at $z = 1.9410, 1.8916, 1.8873, 1.7193$, and 1.1923. 
The SED differs
significantly from system to system, so that we can conclude that the ionization field fluctuates at
the redshift scale of at least $\Delta_z \sim 0.004$. 
This is comparable with $\Delta_z \la 0.01$ revealed from joint observations
of \ion{He}{ii} and \ion{H}{i} Ly-$\alpha$ forest at $2.0 < z < 2.9$ (Shull \etal\ 2004, 2010).  
Figure~\ref{fg16} shows $\eta \equiv N$(\ion{He}{ii})/$N$(\ion{H}{i}), the
parameter used in the forest studies to probe fluctuations of the ionizing background with $z$.
The curves $\eta(U)$ are calculated with CLOUDY for a range of SEDs restored for a given system and
for the gas metallicity [Z] = $-1.0$ (arbitrarily chosen since 
$\eta$-values are very weakly affected by metallicity).
It is seen that while moving from $z=1.9410$ to $z=1.8916$, $\eta$ decreases from $\sim 40$ 
to $\sim 10$ (for $\lg U > -2$ relevant for the majority of the forest absorbers) 
and then increases at $z = 1.8874$  to $\sim 30-50$. 
In the vicinity of $z = 1.7193$, the $\eta$-values reach several hundreds due to a sharp decrease 
in the number of photons with energies $E \ga 4$ Ryd
in the ionizing continuum and, as a consequence, the impossibility to ionize \ion{He}{ii}. 
At $z = 1.1923$, where the preferable SEDs
return to the range detected between $z = 1.9410$ and 1.8873, 
$\eta$ will be again about a few tens. 
The \ion{He}{ii}/\ion{H}{i} forest
studies have shown a slow evolution of $\eta$ toward smaller values at lower redshift, 
with the mean $\bar{\eta} \sim 70$ at $z = 2.9$
and  $\bar{\eta} \sim 40$ at $z = 2.0$ (Zheng \etal\ 2004; Fechner \etal\ 2006). 
Our values are in line with these results.

Further on, the restored continuum shapes allow us to identify 
the main sources contributing to the ionizing background.
At $z = 1.9410, 1.8916, 1.8873$, and 1.1923, these are definitely the AGNs. 
In all systems, simple power-law SEDs are
only marginally consistent with the observed lines, whereas a more appropriate continuum shape 
corresponds to a broken power law with an intensity break of an order of magnitude at 4 Ryd. 
Such an SED is usually attributed to absorption
in the \ion{He}{ii} Lyman continuum occurring when quasar radiation, which is assumed to have 
a simple power-law SED, passes through the IGM
(Haard \& Madau 1996, 2012; Giroux \& Shull 1997). 
However, studying the so-called associated systems 
(absorbers located in the physical vicinity of quasars), 
we found that they were ionized by a continuum that already had
a pronounced break at 4 Ryd (Levshakov \etal\ 2008), i.e., 
\ion{He}{ii} absorption occurred not outside, but within the radiation-generating structures. 
A generally accepted mechanism of the majority of the quasar continuum is photon
emission by geometrically thin but optically thick accretion disc. 
That optically thick structures are indeed involved
is verified observationally by detecting the hydrogen Balmer edge in the spectra of several quasars 
(Kishimoto \etal\ 2003, 2004; Hu \& Zhang 2012). It is
important that the corresponding Lyman break at 1 Ryd ($3.3\times10^{15}$ Hz) is not observed, and
this is just what the  non-LTE models of accretion discs predict 
(Hubeny \etal\ 2000; Blaes \etal\ 2001). The break is reduced and smeared out
both by relativistic effects (Doppler shifts) and 
by summing over \ion{H}{i} emission and absorption edges from different
parts of the disc. 
If the disc is optically thick in \ion{H}{i}, it will be opaque in \ion{He}{ii} as well, and
it is clear that a real depth of the \ion{He}{ii} Lyman break at 4 Ryd in the emerging disc spectrum 
will be affected by the same processes as the \ion{H}{i} Lyman break, primarily by a combination 
of the \ion{He}{ii} emission and absorption.
However, at energies close to and above 4 Ryd, the thermal accretion disc already 
loses its validity as a model for quasar
continuum. At high frequencies (soft and hard X-ray bands) radiation is generated by other, 
still poorly understood, mechanisms;
this means that calculations of the \ion{He}{ii} emission/absorption balance become highly uncertain.
Depending on the parameters, the resulting SEDs demonstrate at $E \ga 4$ Ryd a
variety of shapes, from a sharp flux decrease to an almost complete reduction of a flux discontinuity 
(Hubeny \etal\ 2000, 2001).
As already mentioned in Sect.~\ref{sect-1}, quasar spectra between $\sim 3$ Ryd and the soft X-ray band 
(0.2 keV) are completely absorbed, so that the observational data cannot be used to distinguish 
between the model predictions.
Thereby it seems quite plausible that the integrated spectrum of all radiation-generating quasar structures 
retains a flux break at 4 Ryd.
This break can be further enhanced if the central engine is accompanied by a so-called warm absorber,
a circumnuclear gas responsible for broad absorption features in the X-ray spectra and detected in 
$\sim 50$ \% of AGNs (Piconcelli \etal\ 2005). 
With a total hydrogen column density $N$(H) $\sim 10^{22} - 10^{23}$ \cm,
the ionization parameter $U$ $\sim$~few units, and the standard helium abundance Z = 0.083, 
the predicted (by CLOUDY) column density of \ion{He}{ii}  is $\sim 10^{17} - 10^{18}$ \cm. Hence, 
it noticeably softens the radiation at $E > 4$ Ryd.
Thus, broken power-law SEDs restored for the $z = 1.9410, 1.8916, 1.8873$, and 1.1923 systems  
most probably represent intrinsic AGN spectra as only weakly attenuated, if at all, by the IGM. 

In this regard the differences between SEDs
at $z = 1.9410$ and $z =1.8916$ 
(cf. SEDs in Figs.~\ref{fg5} and \ref{fg10}) can be a consequence of different spectral shapes of the ionizing radiation
from the contributing sources. 
It is well established that both the AGN space density and their cumulative luminosity peak around
$z \sim 2$ (Assef \etal\ 2011, and references therein). 
To make some quantitative estimates, we employ the probability
that a line of sight intersects a `sphere of influence' of an AGN 
in redshift interval $\Delta z$ at redshift $z$ (Peebles 1993):
\begin{equation}
p = \Delta X /\ell_0\ ,
\label{Eq1}
\end{equation}
which is the ratio of the physical path length 
\begin{equation}
\Delta X = \frac{c}{H_0} \frac{(1+z)^2}{\sqrt{\Omega_{\rm M}(1+z)^3 + \Omega_\Lambda}} \Delta z
\label{Eq2}
\end{equation}
and the mean free path between the spheres
\begin{equation}
\ell_0 = \langle n(z) \sigma(z) \rangle^{-1}_0 \ .
\label{Eq3}
\end{equation}
Here $n(z)$ and $\sigma(z)$ are, respectively, the comoving number density of AGNs and cross section of the 
sphere of influence, 
$c$ is the speed of light, $H_0$ is the Hubble constant, and $\Omega_{\rm M}$, $\Omega_\Lambda$ are
dimensionless density parameters 
(we adopt the flat $\Lambda$CDM cosmology with $H_0 = 70$ km~s$^{-1}$~Mpc$^{-1}$, $\Omega_{\rm M} = 0.3$, and
$\Omega_\Lambda = 0.7$).

The cross section of the sphere of influence 
is defined by a radius $\tilde{R}$ at which the AGN radiation flux at 1 Ryd, $J^q_\nu(\tilde{R})$,
becomes equal to the mean background metagalactic flux, $J^{bg}_\nu(z)$.
At redshift $z = 1.9$, the value of $J^{bg}_\nu \sim 6\times10^{-22}$ erg~cm$^{-2}$~s$^{-1}$~Hz$^{-1}$~sr$^{-1}$ at 1 Ryd 
was measured in Tytler \etal\ (2004), Bolton \etal\ (2005), and Dall'Aglio \etal\ (2008).
Taking the AGN comoving space density at this redshift from Wolf \etal\ (2003) and Croom \etal\ (2009) and
choosing $\Delta z = 0.05$, we calculate the intersection probability $p$ 
depending on the source absolute magnitude $M_B$ (blue curve in Fig.~\ref{fg17})\footnote{Conversion of $M_B$ into 
the monochromatic flux at 1 Ryd is performed with the spectral index $\alpha = 0.4$.}.
The red curve in Fig.~\ref{fg17} indicates the comoving AGN emissivity, $\varepsilon$, at 1 Ryd. 
It is seen that bright sources with $M_B < -24$ (quasars) contribute
the majority ($\sim$80\%) of all ionizing photons, and at the same time their spheres of influence 
are the most probable candidates for intersection. 
Since the space number density of such quasars is low 
(at $z \sim 2$, $n < 5$ objects per cube of 100 Mpc side),
only a small number of them contribute to the ionizing background at
every given point along the line of sight. In addition,
the diversity of their intrinsic SEDs just causes the variations in the spectral shape 
of the resulting ionizing radiation. 

The mean free path $\ell_0$ for all objects with $M_B < -21$ is $\ell_0 = 1.5$ Gpc (comoving). 
Thus, the probability of intersecting a sphere of influence  
of at least one quasar in the redshift interval $\Delta z \sim 0.05$ at $z \sim 1.9$ 
($\Delta X = 0.6$ Gpc) is $p \simeq 0.4$ 
(it may be even higher if obscured quasars are taken into account). 
This probability is sufficient to suggest that at least in one of the three
close systems ($z = 1.9410, 1.8916, 1.8873$) the ionizing radiation is dominated by a local source, i.e., 
by a quasar located transverse to the line of sight. 

The estimate of $p$ can be directly tested against observations. Searching for `quasars near quasars', i.e.,
for foreground quasars located within $30'$ from the line of sight to a background quasar, 
Worseck \etal\ (2008) detected 7 QSOs per 18 lines of sight in the range $1.89 < z < 1.94$,
among them one pair of QSOs separated by $\Delta z < 0.05$. We note here that 
in general such pairs are not extremely rare since the survey by Worseck \etal\  
contains many other examples at different redshifts.
Thus, it is not unlikely that local sources affect the mean metagalactic SED even in
more than one system between $z = 1.94$ and 1.89.
 
The spectral shape of the ionizing radiation can also be altered due to absorption 
in the intervening systems, primarily by strong absorbers 
with $N$(\ion{H}{I}) $> 10^{17}$ \cm.
The common assumption is that such systems are directly related to galaxies and
originate in galactic halos at different impact parameters. 
Then the mean free path between the halos is given by (e.g., Burbidge \etal\ 1977):
\begin{equation}
\ell_0  =
[\sigma_\ast \phi_\ast \Gamma(1+s+2\beta)]^{-1} ,
\label{Eq8}
\end{equation}
where $\Gamma(x)$ is the gamma function. 
For the parameters of Eq.(\ref{Eq8}) the following values are taken: 
$\sigma_\ast = \pi R^2_\ast$  with $R_\ast \approx 100$ kpc, $\beta \approx 0.35$ (e.g., Matejek \& Simcoe 2012),
$\phi_\ast \simeq 2\times10^{-3}$ Mpc$^{-3}$ (corresponding to $M^\ast_B \simeq -21.5$ at $z \sim 2$) and 
$s \simeq -1.25$ (Poli \etal\ 2003; Gabasch \etal\ 2004). 
This provides $\ell_0 \simeq 8$ Gpc. Given $\ell_0 = 1.5$ Gpc for AGNs, the probability of
intersecting a galactic halo at an impact parameter $\la 100$ kpc is several times lower 
as compared to the sphere of influence of an AGN. 

Thus, it is more likely that differences in the SEDs between 
the $z = 1.9410$ and 1.8916 systems 
are indeed due to SEDs in QSOs but not due to the effects of
radiative transfer in the IGM. 
We note that this is an additional argument supporting the intrinsic 
nature of the intensity break at 4 Ryd in the outcoming AGN spectrum.

Of course, the above consideration does not exclude that in some cases the 
shape of the ionizing background radiation can be attenuated by the intervening systems.
For example, if we assume that the systems at $z = 1.8916$ and
$z = 1.8876$ are illuminated by the same transverse quasar, then the softer SED in the latter case 
may indicate the presence of a strong absorber 
with $N$(\ion{H}{I}) $> 5\times10^{16}$ \cm\ between this system and the quasar.

As for the SED at $z = 1.7193$, 
the footprint of a local source is quite certain: a deep intensity drop at $E > 3.5$ Ryd
points to a substantial input from a
soft (galactic, i.e., stellar) component (see, e.g., Fig.~1 in Giroux \& Shull 1997). 
In a quasar-dominated field of metagalactic ionizing radiation, which is relatively hard at $E > 4$ Ryd,
such spectral shapes can be sensed only in the vicinity of the source itself.  
Possible emitters of radiation with this type of SED are the so-called (ultra) 
luminous infrared galaxies, (U)LIRGs, which in many cases show spectra powered
by a mixture of AGN and starburst activity (Sanders \etal\ 1988). 
The (U)LIRGs that host AGNs also reveal large-scale
gas outflows (Alexander \etal\ 2010; Harrison \etal\ 2012). At low $z$ ($\sim 0.05$), 
it is even possible to resolve multiple star clusters with a young stellar population 
(Mazzarella \etal\ 2012) and superbubbles in the blowout phase, which
are believed to form as a result of giant supernova explosions (Lipari \etal\ 2005, 2009).
In addition, the $z = 1.7193$ absorber could be an outflow from a transverse (U)LIRG
because the overabundance of $\alpha$-elements and general low metallicity of the gas are typical
for the enrichment by massive star(s) exploded as core-collapse supernova (SN II).

The difference in the high-frequency range of the SED between pure AGN and AGN+starburst sources 
is illustrated in Fig.~\ref{fg18}, which shows observations by Grupe \etal\ (2010) 
that were taken with SWIFT simultaneously in the optical/UV 
and soft X-ray bands. Blue stars indicate data for \object{Mkn~335}, which is a classical AGN with 
no star-forming activity, and red stars show
data for \object{Mkn~876}, a galaxy with both strong AGN
activity and intense circumnuclear star formation (Shi \etal\ 2007; Sani \etal\ 2010). 
The latter SED is appreciably softer in the X-ray band. 
For comparison, SEDs restored for the $z = 1.9410$ and 1.7193 systems are also shown by dotted lines. 
It is seen that they fit well in the gap between the UV and X-ray points.

\section{Summary}
\label{sect-6}

The spectrum of bright quasar \object{HS~1103+6416} (\zem\ = 2.19) 
available in the wavelength coverage from 1400 \AA\ to 5700 \AA\ contains
between the redshifts $z = 1.19$ and 1.94 
five successive absorption systems detached by the redshifts intervals $\Delta_z \sim 0.004 - 0.5$.
These systems are optically thin both in \ion{H}{i} and in \ion{He}{ii} 
and reveal multiple lines of different metal ions,
the  ionization potentials of which sample the EUV range (1 Ryd -- 0.2 keV)
of the ionizing radiation with a fairly small step. Such properties make it possible to solve with sufficient
accuracy the inverse problem of spectroscopy: namely, to estimate the physical 
parameters of the absorbing medium simultaneously
with the reconstruction of the SED in the underlying ionizing continuum. 
We note that metal absorption systems in quasar spectra are the only means to probe observationally the 
SED of the ionizing radiation at redshifts
$z \la 2$. The results obtained are as follow.
\begin{enumerate}
\item[1.]
The SED changes significantly from one system to another, i.e., the intergalactic ionization field
fluctuates at the scale of at least $\Delta_z \sim 0.004$. This is consistent with $\Delta_z \la 0.01$ estimated
from \ion{He}{ii} and \ion{H}{i} Lyman-$\alpha$ forest measurements between redshifts 2 and 3.
\item[2.]
In all systems, pure power law SEDs of the type $J_\nu \propto \nu^{-\alpha}$ are not or only marginally consistent 
with the observed line profiles.
A more preferable continuum shape is a broken power-law with the break at $E ~ 4$ Ryd,
which likely represents an intrinsic SED of the outcoming AGN radiation. 
\item[3.]
The SEDs restored for 
the \zabs\ = 1.1923, 1.8873, 1.8916, and 1.9410 systems
point to AGNs/QSOs as the main source of the ionizing radiation. 
The SED variability is mostly caused by a small number of QSOs contributing
at any given redshift to the ionizing background. 
At scales $\Delta_z \ga 0.05$ the influence of local sources (foreground quasars located transverse to the
line of sight) becomes significant. 
This can be tested observationally by a search for QSOs close to the line of
sight as done by Worseck \etal\ (2008).
\item[4.] 
A remarkable type of SED is restored for the \zabs\ = 1.7193 system. 
It demonstrates a sharp break shifted to $E \sim 3.5$ Ryd with
a subsequent intensity decrease by $\sim$1.5 dex, suggesting 
comparable inputs from both hard (AGNs/QSOs) and soft (young stars) 
radiation components. In a quasar-dominated ionizing radiation field, such a soft continuum
can survive only in the vicinity of the radiation source itself. 
For the given system
this source is probably one of the (U)LIRGs, many of which reveal both strong AGN
activity and intense star formation in the circumnuclear regions. 

\end{enumerate}

\begin{acknowledgements}
IIA and SAL are grateful for the kind hospitality of
Hamburger Sternwarte, where this work has been done.
We also thank an anonymous referee for the useful comments received. 
The work of IIA and SAL is supported by
DFG Sonderforschungsbereich SFB 676 Teilprojekt C4
and, in part, by
the State Program `Leading Scientific Schools of Russian Federation'
(grant NSh 4035.2012.2).
\end{acknowledgements}

\clearpage

\clearpage

\clearpage

\clearpage

\clearpage

\clearpage

\clearpage

\clearpage

\clearpage

\clearpage

\clearpage

\clearpage

\clearpage

\clearpage

\clearpage
\begin{figure}
\vspace{0.0cm}
\hspace{-2.2cm}\psfig{figure=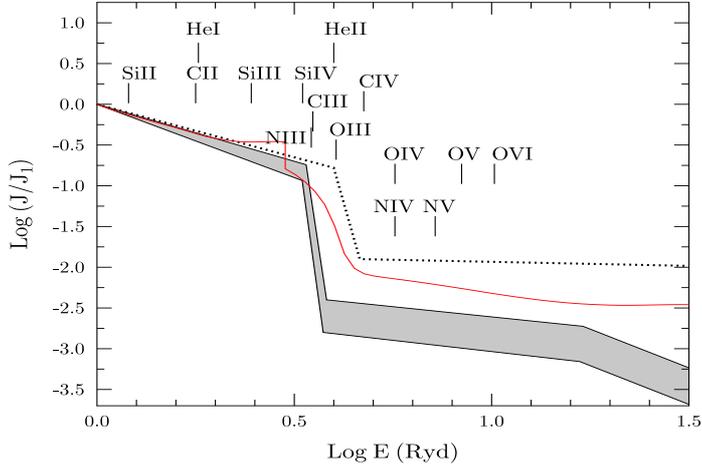,height=15cm,width=13cm}
\vspace{-4.0cm}
\caption[]{
Range of SEDs of the ionizing UV radiation (shadowed area) reconstructed from absorption
lines at \zabs\ = 1.7193. The initial guess SED is plotted by the dotted line.
For comparison, the
model metagalactic SED of Haardt \& Madau (1996)
is shown by red.
}
\label{fg15}
\end{figure}

\clearpage
\begin{figure}
\vspace{0.0cm}
\hspace{-2.2cm}\psfig{figure=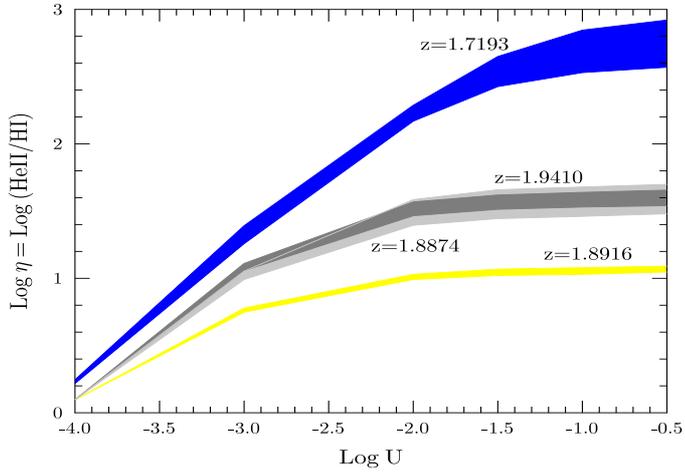,height=15cm,width=13cm}
\vspace{-4.0cm}
\caption[]{
The parameter $\eta$ = \ion{He}{ii}/\ion{H}{i} as a function of the ionization parameter $U$
calculated for the SEDs restored at $z = 1.9410$ (dark grey), 
1.8916 (yellow), 1.8874 (light grey), and 1.7193 (blue). 
}
\label{fg16}
\end{figure}

\clearpage
\begin{figure}
\vspace{0.0cm}
\hspace{-1.0cm}\psfig{figure=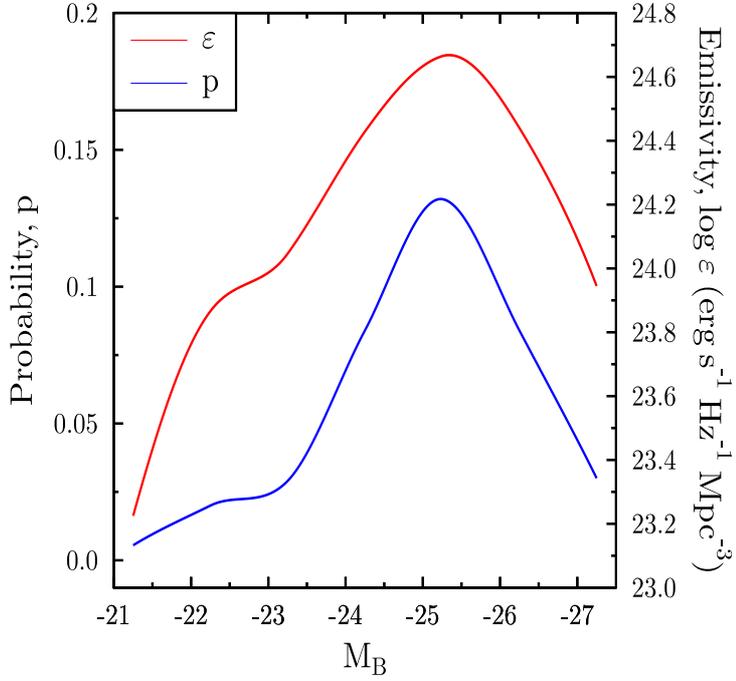,height=15cm,width=13cm}
\vspace{-1.0cm}
\caption[]{
Blue curve: the probability for a given line of sight to intersect the sphere of influence of an AGN with 
an absolute magnitude $M_B$; 
red curve: comoving AGN emissivity at 1 Ryd. Both curves are calculated for $z = 1.9$. 
}
\label{fg17}
\end{figure}

\clearpage
\begin{figure}
\vspace{0.0cm}
\hspace{-2.2cm}\psfig{figure=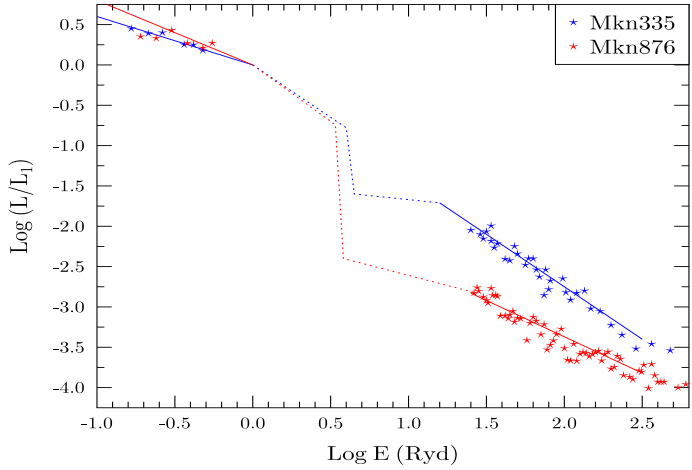,height=15cm,width=13cm}
\vspace{-4.0cm}
\caption[]{
Simultaneous UV and soft X-ray SWIFT observations by Grupe \etal\ (2010) of two AGN sources
\object{Mkn~335} and \object{Mkn~876} (blue and red stars, respectively).
\object{Mkn~335} is a pure AGN
without significant star-formation activity, whereas \object{Mkn~876} demonstrates both strong
AGN activity and intense circumnuclear star formation. 
The original data by Grupe \etal\ are ($i$) corrected for Galactic absorption using
the X-ray cross section from Morrison \& McCammon (1983) and
$N$(H) from Table~1 in Grupe \etal\ ; ($ii$) transferred from coordinates 
$\log (\nu L_\nu)$ vs. $\log \nu$ to $\log (L_\nu)$ vs. $\log E$;
($iii$) normalized by $L_\nu$ at 1 Ryd, which is found by linear extrapolation of the UV data to 1 Ryd
with spectral indices reported for \object{Mkn~335} and \object{Mkn~876} in Shull \etal\ (2012). 
The solid lines in the X-ray range
are the power laws with indices given in Table~4 in Grupe \etal . The dotted lines show the SEDs restored
for the $z = 1.9410$ (blue) and 1.7193 (red) systems.
}
\label{fg18}
\end{figure}

\clearpage
\begin{table*}[t!]
\centering
\caption{Calculated column densities (in \cm) for atoms and ions in systems used
to reconstruct the SED of the underlying ionizing continuum
}
\label{tbl-1}
\begin{tabular}{ccccccc}
\hline
\hline
\noalign{\smallskip}
H\,{\sc i} & \hspace{-0.1cm}C\,{\sc ii} & Si\,{\sc ii} & \hspace{-0.2cm}O\,{\sc i} &
N\,{\sc iv} & \hspace{-0.1cm}Al\,{\sc ii} & \hspace{-0.1cm}Fe\,{\sc ii}\\[-2pt]
 &  C\,{\sc iii} & \hspace{0.1cm}Si\,{\sc iii} & \hspace{-0.2cm}O\,{\sc iii} & 
\hspace{-0.1cm}N\,{\sc v} & Al\,{\sc iii} \\[-2pt]
 &  C\,{\sc iv} & \hspace{0.1cm}Si\,{\sc iv} & \hspace{-0.2cm}O\,{\sc iv}\\
\noalign{\smallskip}
\hline
\noalign{\medskip}
\multicolumn{7}{c}{\it \zabs\ = 1.9410}\\
\noalign{\smallskip}
(2.5--3.0)E15 & (5.6$\pm$0.6)E12 & (3.5$\pm$0.5)E11 & (6.3--7.8)E14$^b$ & $\ldots$ & $\ldots$ & $\ldots$ \\[-1pt]
              & 2.6E14$^a$   & (8.4$\pm$0.9)E12 & (1.1--1.4)1E15$^b$ & $<$1.5E13 & $\ldots$ \\[-1pt]
              & (2.2$\pm$0.1)E14 & (1.4$\pm$0.2)E13 \\[-1pt]
\noalign{\medskip}
\multicolumn{7}{c}{\it \zabs\ = 1.8916A}\\
\noalign{\smallskip}
(3.5--4.5)E16 & (5.2$\pm$0.2)E13 & (8.8$\pm$0.5)E12 & (9.3$\pm$0.3)E12 & $\ldots$ 
& (7.5$\pm$0.8)E11 & (2.7$\pm$0.3)E12 \\[-1pt]  
              & $\ldots$              & 9.1E13 & $\ldots$ & $\ldots$ & 3.5E11 \\[-1pt]
              &(3.8$\pm$0.7)E12  & (2.5$\pm$0.3)E12 \\[-1pt]
\noalign{\medskip}
\multicolumn{7}{c}{\it \zabs\ = 1.8873}\\
\noalign{\smallskip}
(4.8--5.2)E15 & $<$3.5E12 & $\ldots$ & $\ldots$ & $<$3.0E13 & $\ldots$ & $\ldots$ \\[-1pt]
              & 9.0E13$^a$ & (2.2$\pm$0.4)E12  & (1.8--2.2)E14$^b$ & $<$6.0E12 & $\ldots$ \\[-1pt]
              & (7.3$\pm$0.2)E13 & (2.2$\pm$0.2)E12 & (3.9--4.9)E14$^b$ \\[-1pt]
\noalign{\medskip}
\multicolumn{7}{c}{\it \zabs\ = 1.7193}\\
\noalign{\smallskip}
(1.4--1.5)E16 & 2.5E12$^a$ & $<$4.0E11 & $\ldots$ & $\ldots$ & $\ldots$ & $\ldots$ \\[-1pt]
              & 1.0E14$^a$ & (9.8$\pm$0.2)E12 & 4.6E14$^c$ & $\ldots$ & $\ldots$ \\[-1pt]
              & (1.06$\pm$0.05)E13 & (4.7$\pm$0.3)E12 & 2.5E14$^c$\\[-1pt]
\noalign{\medskip}
\multicolumn{7}{c}{\it \zabs\ = 1.1923}\\
\noalign{\smallskip}
$\sim 5$E15 & $\ldots$ & $\ldots$ & $\ldots$ & $\ldots$ & $\ldots$ & $\ldots$ \\[-1pt]
            & (1.5--2.0)E14 & $\ldots$ & (4.0--5.0)E14 & $\ldots$ & $\ldots$\\[-1pt]
            & (2.2$\pm$0.2)E14 & $\ldots$ & (1.0--1.5)E15 \\[-1pt]
\noalign{\smallskip}
\hline
\noalign{\smallskip}
\multicolumn{7}{l}{{\bf Notes.} $^a$Calculated value with C content and density-velocity 
distribution derived from high-resolution}\\[-2pt]
\multicolumn{7}{l}{metal lines .  
$^b$Calculated value assuming [O/C] = 0.2-0.3. $^c$Calculated value assuming [O/C] = 0.5  }\\
\end{tabular}
\end{table*}

\begin{table*}[t!]
\centering
\caption{Mean ionization parameters, $U_0$, and relative element
abundances, [X/Y], for the systems from Table~\ref{tbl-1}.
[X/Y] =  $\log (N_{\rm X}/N_{\rm Y}) - \log (N_{\rm X}/N_{\rm Y})_\odot$.
}
\label{tbl-2}
\begin{tabular}{lccccccc}
\hline
\hline
\noalign{\smallskip}
\multicolumn{1}{c}{\zabs} & $\log U_0$ & [C/H] & [Si/C] & [O/C] & [N/C] & [Al/C] & [Fe/C]\\
\noalign{\smallskip}
\hline
\noalign{\medskip}
1.9410 & -2.1 -- -2.0 & -0.3 -- -0.4 & 0.0 -- 0.1 & 0.2 -- 0.3 &  $<$0.0 & $\ldots$ \\[-1pt]
1.8916A& -3.3 -- -3.2 & -0.4 -- -0.5 & 0.0 & 0.2 & $\ldots$ & 0.1 -- 0.2 & 0.0 \\[-1pt]
1.8873 & -2.1 -- -2.0 & -1.1 -- -1.2 & 0.0 -- 0.2 & 0.2 -- 0.3 & $\leq$0.0 & $\ldots$\\[-1pt]
1.7193 & -2.0 -- -1.75 & -2.2 -- -2.3 & 0.0 -- 0.3 & $<$0.5 & $\ldots$ & $\ldots$ \\[-1pt]
1.1923 & -1.9 -- -1.7 & -1.2 -- -1.5 & $\ldots$ & 0.2 -- 0.3 & $\ldots$ & $\ldots$ \\[-1pt]
\noalign{\smallskip}
\hline
\end{tabular}
\end{table*}

\begin{table*}[t!]
\centering
\caption{Deconvolution parameters of the \ion{H}{i} profiles in the $z = 1.8916$ system (Fig.~\ref{fg9})
}
\label{tbl-3}
\begin{tabular}{rccrl}
\hline
\hline
\noalign{\smallskip}
\multicolumn{1}{c}{Component} & Central & Broadening & \multicolumn{1}{c}{Column} & Comments \\[-2pt]
\multicolumn{1}{c}{No.} & velocity $V$, \kms & parameter $b$, \kms & \multicolumn{1}{c}{density $N$, \cm} \\
\noalign{\smallskip}
\hline
\noalign{\medskip}
\multicolumn{1}{c}{1} & $-446.0$ & 30.0 & (4.8--5.2)E15 & system $z=1.8873$ \\[-1pt]
\multicolumn{1}{c}{2} & $-364.0$ & 34.0 & 1.4E14 \\[-1pt]
\multicolumn{1}{c}{3} & $-126.0$ & 36.0 & 9.0E13 \\[-1pt]
\multicolumn{1}{c}{4} & $-91.0$ & 14.0 & (3.0--5.0)E14 \\[-1pt]
\multicolumn{1}{c}{5} & $-55.4$ & 17.0 & (3.5--4.5)E16 & system $A$, Fig.~\ref{fg8} \\[-1pt]
\multicolumn{1}{c}{6} & 9.5 & 17.0 & (1.0--1.1)E17 & system $B$, Fig.~\ref{fg8} \\[-1pt]
\multicolumn{1}{c}{7} & 30.5 & 15.0 & (4.5--5.0)E16 & system $C$, Fig.~\ref{fg8} \\[-1pt]
\multicolumn{1}{c}{8} & 68.0 & 19.0 & (8.5--9.5)E16 & system $D$, Fig.~\ref{fg8} \\[-1pt]
\multicolumn{1}{c}{9} & 127.5 & 25.0 & 1.4E14 \\[-1pt]
\multicolumn{1}{c}{10} & 271.0 & 29.0 & 2.2E14 \\
\noalign{\smallskip}
\hline
\end{tabular}
\end{table*}


\begin{thebibliography}{}
\bibitem{}Agafonova, I. I., Molaro, P., Levshakov, S. A., \& Hou, J. L. 2011, A\&A, 529, A28 

\bibitem{}Agafonova, I. I., Centuri\'on, M., Levshakov, S. A., \& Molaro, P. 2005, A\&A, 441, 9

\bibitem{}Akerman, C. J., Carigi, L., Nissen, P. E., Pettini, M., \& Asplund, M. 2004, A\&A, 414, 93

\bibitem{}Alexander, D. M., Swinbank, A. M., Smail, I., McDermid, R., Nesvalda, N. P. H. 2010, MNRAS, 402, 2211

\bibitem{}Asplund, M., Grevesse, N., Sauval, A. J. \& Scott, P. 2009, ARA\&A, 47, 481

\bibitem{}Assef, R. J., Kochanek, C. S., Ashby, M. L. N., Brodwin, M., Cool, R. \etal\ 2011, ApJ, 728, 56

\bibitem{}Bensby, T., \& Feltzing, S. 2006, MNRAS, 367, 1181

\bibitem{}Blaes, O., Hubeny, I., Agol, E., \& Krolik, J. H. 2001, ApJ, 563, 560

\bibitem{}Bresolin, F. 2007, ApJ, 656, 186

\bibitem{}Bolton, J.S., Haehnelt, M.G., Viel, M., \& Springel, V. 2005, MNRAS, 357, 1178

\bibitem{}Burbidge, G., O'Dell, S. L., Roberts, D. H., Smith, H. E. 1977, ApJ, 218, 33

\bibitem{}Cooke, R., Pettini, M., Steidel, C. C., Rudie, G. C., \& Nissen, P. E. 2011, MNRAS, 417, 1534

\bibitem{}Croom, S.M., Richards, G.T., Shanks, T., Boyle, B.J. \etal\ 2009, MNRAS, 399, 1755

\bibitem{}Dall'Aglio, A., Wisotzki, L., \& Worseck, G. 2008, A\&A, 491, 465

\bibitem{}Erni, P., Richter, P., Ledoux, C., \& Petitjean, P. 2006, A\&A, 451, 19

\bibitem{}Esteban, C., Bresoli, F., Peimbert, M., Garcia-Rojas, J., Peimbert, A., \& Mesa-Delgado, A. 2009, ApJ, 700, 654

\bibitem{}Fardal, M. A., Giroux, M. L., \& Shull, J. M. 1998, AJ, 115, 2206

\bibitem{}Fechner, C. 2011, A\&A, 532, 62

\bibitem{}Fechner, C., \& Reimers, D. 2007, A\&A, 461, 847

\bibitem{}Fechner, C., Reimers, D., Songaila, A., Simcoe, R. A. \etal\ 2006, A\&A, 455, 73

\bibitem{}Ferland, G. J., Korista, K. T., Verner, D. A., \etal\ 1998, PASP, 110, 761

\bibitem{}Gabasch, A., Bender, R., Seitz, S. \etal\ 2004, A\&A, 421, 41

\bibitem{}Garcia-Rojas, J., Esteban, C., Peimbert, M., Rodrigez, M., Ruiz, M. T. \& Peimbert, A. 2004, ApJS, 153, 501

\bibitem{}Garnett, D. R., Dufour, R. J., Peimbert, M. \etal\ 1995, ApJL, 449, L77

\bibitem{}Giroux, M. L., \& Shull, J. M. 1997, AJ, 113, 1505

\bibitem{}Grevesse, N., Asplund, M., Sauval, A. J., \& Scott, P. 2010, Ap\&SS, 328, 179

\bibitem{}Griest, K., Whitmore, J. B., Wolfe, A. M., \etal\ 2010, ApJ, 708, 158

\bibitem{}Grupe, D., Komossa, S., Leighly, K. M. \& Page, K. L. 2010, ApJS, 187, 64

\bibitem{}Haard, F., \& Madau, P. 2012, ApJ, 746, 125

\bibitem{}Haard, F., \& Madau, P. 1996, ApJ, 461, 20

\bibitem{}Harrison, C. M., Alexander, D. M., Swinbank, A. M. \etal\ 2012, MNRAS, 426, 1073

\bibitem{}Hu, R., \& Zhang, S.-N. 2012, MNRAS, 426, 2847

\bibitem{}Hubeny, I., Blaes, O., Krolik, J. H. \& Agol, E. 2001, ApJ, 559, 680

\bibitem{}Hubeny, I., Agol, E., Blaes, O., \& Krolik, J. H. 2000, ApJ, 533, 710

\bibitem{}Jakobsen, P., Jansen, R. A., Wagner, S., Reimers, D. 2003, A\&A, 397, 891

\bibitem{}Kingsburgh, R. L., \& Barlow, M. J. 1999, MNRAS, 271, 257

\bibitem{}Kishimoto, M., Antonucci, R., Boisson, C., \& Blaes, O. 2004, MNRAS, 354, 1065

\bibitem{}Kishimoto, M., Antonucci, R., \& Blaes, O. 2003, MNRAS, 345, 253

\bibitem{}K\"ohler, S., Reimers, D., Tytler, D., \etal\ 1999, A\&A, 342, 395

\bibitem{}Levshakov, S. A., Agafonova, I. I., Molaro, P., Reimers, D., \& Hou, J. 2009, A\&A, 507, 209

\bibitem{}Levshakov, S. A., Agafonova, I. I., Reimers, D., Hou, J. \etal\ 2008, A\&A, 483, 19

\bibitem{}Levshakov, S. A., Agafonova, I. I., \& Kegel, W. H. 2000, A\&A, 360, 833

\bibitem{}Lipari, S., Sanchez, S. F., Bergmann, M. \etal\ 2009, MNRAS, 392, 1295

\bibitem{}Lipari, S., Terlevich, R., Zheng, W. \etal\ 2005, MNRAS, 360, 416

\bibitem{}Matejek, M. S., \& Simcoe, R. A. 2012, ApJ, 761, 112

\bibitem{}Morrison, R., \& McCammon, D. 1983, ApJ, 270, 119

\bibitem{}Morton, D. C. 2003, ApJS, 149, 205

\bibitem{}Peebles P. J. E. 1993, {\it Principles of Physical Cosmology} (Princeton University Press, Princeton)

\bibitem{}Piconcelli, E., Jimenez-Bailon, E., Guainazzi, M., \etal\ 2005, A\&A, 432, 15

\bibitem{}Poli, F., Giallongo, E., Fontana, A., \etal\ 2003, ApJ, 593, L1

\bibitem{}Reimers, D., Kohler, S., Wisotzki, L., Groote, D., Rodriguez-Pascual, P., \& Wamsteker, W.
1997, A\&A, 327, 890

\bibitem{}Reimers, D., Rodriguez-Pascual, P., Hagen, H.-J., \& Wisotzki, L. 1995, A\&A, 293, L21

\bibitem{}Samson, J. A. R., He, Z. X., Yin, L., \& Haddad, G. N. 1994, JPhysB, 27, 887 

\bibitem{}Sanders, D. B., Soifer, B. T., Elias, J. H., Neugebauer, G., \& Matthews, K. 1988, ApJL, 328, 35

\bibitem{}Sani, E., Lutz, D., Risaliti, G., Netzer, H. \etal\ 2010, MNRAS, 403, 1246

\bibitem{}Scott, j.e., Kriss, G.A., Brotherton, M. \etal\ 2004, ApJ, 615, 135

\bibitem{}Shi, Y., Ogle, P., Rieke, G., Antonucci, R. \etal\ 2007, ApJ, 669, 841

\bibitem{}Shull, J. M., Stevans, M., \& Danforth, C. W. 2012, ApJ, 752, 162

\bibitem{}Shull, M. J., France, K., Danforth, C., Smith, B., \& Tumlison, J. 2010, ApJ, 722, 1312

\bibitem{}Shull, J. M., Tumlison, J., Giroux, M. L., Kriss, G. A., \& Reimers, D. 2004, ApJ, 600, 768

\bibitem{}Shull, M.J., Stevans, M., \& Danforth, C. 2012, ApJ, 752, 162

\bibitem{}Tytler, D., Kirkman, D., O'Meara, J.M., \etal\ 2004, ApJ, 617, 1

\bibitem{}Verner, D. A., Barthel, P. D., \& Tytler, D. 1994, A\&AS, 108, 287

\bibitem{}Vogt S. S., Allen S. L., Bigelow B. C., \etal\ 1994, Proc. SPIE, 2198, 362

\bibitem{}Wolf, C., Wisotzki, L., Borch, A., \etal\ 2003, A\&A, 408, 499

\bibitem{}Worseck, G., Fechner, C., Wisotzki, L., Dall'Aglio, A. 2007, A\&A, 473, 805

\bibitem{}Zheng, W., Kriss, G.A., Deharveng, J.-M., Dixon, J. W. \etal\ 2004, ApJ, 605, 631

\end{thebibliography}
\end{document}